\documentclass[twocolumn,preprintnumbers,superscriptaddress]{revtex4}
\usepackage{mathrsfs,amsmath,amsthm,amssymb,amsfonts,graphicx,xcolor}

\newtheorem{definition}{Definition}
\newtheorem{proposition}{Proposition}
\newtheorem{theorem}{Theorem}
\usepackage{makecell}
\usepackage{graphicx}
\usepackage{subcaption}
\usepackage{mathtools}

\usepackage[colorlinks=true,
            linkcolor=black,
            anchorcolor=black,
            citecolor=black,
            urlcolor=blue
            ]{hyperref}

\begin{document}

\title{Mutual Mana: Converting Local Magic into Correlations via Discrete Beamsplitters}
\author{Linshuai Zhang}
\affiliation{School of Mathematics and Computational Science, Xiangtan University, Xiangtan 411105, China}
\author{Huihui Li}
\email{lihh@mail.tsinghua.edu.cn}
\affiliation{Yau Mathematical Sciences Center, Tsinghua University, Beijing 100084, China}

\begin{abstract}



Magic (non-stabilizerness) is a key resource for achieving universal  fault-tolerant quantum computation beyond classical computation. 
While previous studies have primarily focused on magic in single systems, its interactions and distribution in multipartite settings remain largely unexplored.
In this work, we introduce mutual mana as a measure of magic correlations  defined in close analogy with quantum mutual information. Our definition builds upon mana, which is the established quantifier of magic based on discrete Wigner function negativity. We characterize  magic correlations generated by discrete beamsplitters, whose Gaussian counterparts are fundamental components  in quantum optics and quantum  technologies.
We show that coupling  a magic state with a stabilizer vacuum state  via  a discrete beamsplitter  will induce a full conversion of local magic into mutual mana, thereby establishing a mechanism for redistributing magic resources as  magic correlations.  We reveal the  fundamental properties  of mutual mana and derive its explicit expressions   for several prototypical  qutrit states  subject to a discrete beamsplitter. We make a comparative  study of mutual mana with  several  established quantifiers of  correlations generated  by  the qutrit beamsplitter, including quantum mutual information,  mutual $L^1$-norm magic, and mutual stabilizer 2-Rényi entropy.


\end{abstract}

\maketitle

\section{Introduction}

Quantum computation leveraging non-stabilizer resources holds the potential for enormous speedups in a range of quantum protocols and tasks such as quantum teleportation \cite{Gottesman1999}, quantum simulation \cite{BravyiGosset2016,seddon2021}, quantum error correction \cite{liuzw2022,weif2025}, and  quantum many-body physics \cite{chenj2024,WEIFLIU2024}.  
The Gottesman–Knill theorem establishes that quantum circuits composed solely of stabilizer states and Clifford operations can be efficiently simulated classically, and therefore cannot exhibit  any quantum computational advantage \cite{Gottesman1997,Gottesman1998,Gottesman1999,Aaronson2004}. 
 The additional resource that elevates stabilizer computation to universality is known as magic, or nonstabilizerness, which quantifies the deviation of a quantum state from the stabilizer polytope \cite{liuzw2022}. Magic plays a central role in fault-tolerant schemes, where non-Clifford gates are implemented via state injection of magic states \cite{BravyiHaah2012}.

A variety of monotones have been proposed to quantify magic, including robustness of magic \cite{Howard2017}, thauma \cite{WangWilde2020}, relative entropy of magic \cite{Regula2017,ziwenliu2019}, stabilizer rank  \cite{BravyiSmith2016,BravyiBrowne2019}, stabilizer extent \cite{BravyiBrowne2019},  $L^p$-norm magic of  characteristic function \cite{Dai2022,feng2022}, and the mana derived from discrete Wigner-function negativity \cite{Veitch2014}, etc. Among these, mana is distinguished by its computability and additivity. 
Experimental and numerical protocols for detecting and quantifying magic have also been developed, highlighting both its fundamental and practical importance \cite{Souza2011,Rodriguez2025}.
While extensive research has  focused on magic in individual quantum systems, recent efforts have extended into many-body   systems \cite{liuzw2022,Cao2024,CCao2024,weif2025,Andreadakis2025,Ding2025}. 
A notable advance in this context is the use of magic as a tool for quantifying  correlations in composite systems \cite{WhiteCao2021,Leone2022,Frau2024,Tarabunga2025,Frau2025,Qian2025,Feng2025}.
Magic correlations exhibit substantially different dynamical properties and operational signatures compared to conventional quantum correlations such as entanglement \cite{Horodecki2009,Nielsen2010}, quantum nonlocality \cite{Aspect1982,Rowe2001}, and quantum discord \cite{Ollivier2001,Henderson2001,SLUO2008}. Their study thus continues to provide profound insights into the interactive behavior of quantum resources. Nevertheless, most measures of magic correlations remain computationally challenging, even with numerical methods. This motivates the development of alternative quantification approaches grounded in different theoretical perspectives, along with further investigation  into their physical implications and experimental applications.



 Beamsplitters are essential devices in numerous pioneering experiments  in the fields of optics \cite{Loudon2000}, quantum mechanics \cite{Fearn1987,Campos1989,Weihs2001,Sasada2003,Yangchang2024}, and relativity theory \cite{Lammerzahl1999,Borde2004}, such as  Bell test experiments \cite{Brendel1992,Marinkovic2018} and interferometric studies \cite{Zetie2000,Berrada2013}. In quantum information science, they provide a versatile platform for generating entanglement \cite{XWang2002}, enabling universal linear optical quantum computing \cite{Chakhmakhchyan2018} as well as implementing continuous-variable quantum information processing \cite{kim2002,Qureshi2018,fu2020,liluoyue2025}.
Recent advances have extended   beamsplitters to discrete-variable quantum  systems \cite{kbu2023,KBU22023}. Notably, fundamental quantum gates such as the CNOT and swap gates emerge as particular instances of  discrete   beamsplitters.  Nonclassicality and coherence act as fundamental prerequisites for generating quantum correlations with beamsplitters in continuous-variable and discrete-variable quantum systems, respectively \cite{kim2002,XWang2002,OlivaresParis2011,Brunelli2015,SFuLuo2019,Feng2025}.  
We evaluate the magic correlations present in the output states generated by passing product input states through discrete beamsplitters.


In this work, we introduce the notion of mutual mana as a quantifier of  magic correlations  in bipartite quantum systems. Mutual mana vanishes for product states and remains invariant under local Clifford operations.
 We further explore the creation of mutual mana via discrete   beamsplitters. We establish that the interaction between a magic state and a stabilizer vacuum through discrete   beamsplitters results in the complete conversion of input local mana into output mutual mana. This behavior is deeply rooted in the fundamental nature of discrete   beamsplitters as Clifford operators, which conserve the overall magic resource while transforming its distribution across subsystems. The observed resource conversion exemplifies a fundamental principle in magic resource theory: Clifford operations can redistribute magic without creating or destroying it. Consequently, this process offers a conceptually clear and technically feasible method for engineering magic correlations in quantum devices.


The remainder of this paper is organized as follows. In Section II, we  review  the discrete Heisenberg Heisenberg-Weyl group, stabilizer
formalism,  Wigner function and the definition of mana.  We  establish an upper bound for mana  based on  the  purity of quantum states.   We then   briefly review  the discrete  beamsplitters and their  basic properties. 
In Section III, we  introduce   the concept of mutual mana as a measure of magic correlations  and reveal its  fundamental  properties.   We evaluate  mutual mana generated  by discrete   beamsplitters for several  representative  qutrit states.
  In Section IV, we present a comparative analysis between mutual mana and other quantifiers of magic correlations, including mutual stabilizer $\alpha$-Rényi entropy and mutual $L^1$-norm magic. 
In Section V, we conclude with a summary and discussion.  Detailed mathematical proofs  are provided in the Appendix.



\section{Preliminaries}
In this section, we begin with a review of the discrete Heisenberg–Weyl group and the stabilizer formalism  \cite{Gottesman1997,Gottesman1998}. We then introduce the discrete Wigner function in prime power dimensions \cite{Wootters1987,Gibbons2004} and a well-established quantifier of magic known as mana, defined via the discrete Wigner function  negativity \cite{Veitch2014}. Furthermore, we establish an upper bound for mana in terms of the purity of quantum states. Finally, we present an overview of discrete beamsplitters and their fundamental properties \cite{kbu2023,KBU22023,Feng2025}.

\subsection{Discrete Heisenberg--Weyl group and stabilizer formalism}

For a natural number $d$, let $\mathbb{Z}_d$ denote the ring of integers modulo $d$ and 
$\mathbb{C}^d$ the corresponding $d$-dimensional Hilbert space with computational basis $\{|j\rangle:j\in\mathbb{Z}_d\}$. 
We focus on odd prime dimensions $d$, where, according to the discrete Hudson theorem~\cite{Gross2006,Veitch2012}, stabilizer pure states are exactly those with nonnegative discrete Wigner functions, and the mana defined via the negativity of the discrete Wigner function, is a faithful resource measure of nonstabilizerness for pure magic states.

The shift operator $X$ and the phase operator $Z$ are defined as
\begin{equation*}
    X = \sum_{j \in \mathbb{Z}_d} |j+1\rangle\!\langle j|, 
    \qquad 
    Z = \sum_{j \in \mathbb{Z}_d} \omega^j |j\rangle\!\langle j|,
\end{equation*}
where $\omega = e^{2\pi {\rm i}/d}$ is the $d$-th root of unity and addition in the index is taken modulo $d$.
They are unitary, satisfy $X^d = Z^d = \mathbf{1}$, and obey the canonical commutation relation $ZX = \omega XZ$.

For $k,l\in\mathbb{Z}_d$, the corresponding discrete Heisenberg--Weyl operator $D_{k,l}$ is defined as
\begin{equation*}
    D_{k,l} = \tau^{kl} X^{k} Z^{l},
\end{equation*}
where $\tau = -e^{\pi{\rm i}/d}$.
These operators are unitary and satisfy the multiplication relation
\begin{equation*}
    D_{k,l} D_{s,t} = \tau^{\,ls - kt} D_{k+s,\,l+t},\qquad k,l,s,t \in \mathbb{Z}_d.
\end{equation*}
The  generalized Pauli group (discrete Heisenberg--Weyl group) is defined as
\begin{align*}
    \mathcal{P} = \{\tau^j D_{k,l} : j \in \mathbb{Z},\ k,l \in \mathbb{Z}_d\}.
\end{align*}
For an $n$-qudit Hilbert space $(\mathbb{C}^d)^{\otimes n}$, the generalized Pauli group  ($n$-qudit discrete Heisenberg-Weyl group) is defined as 
\begin{equation}\label{gpg}
    \mathcal{P}_n = \mathcal{P}^{\otimes n},
\end{equation}
The Clifford group is defined as the normalizer of the Pauli group in the full unitary group,
\begin{align*}
    \mathcal{C} &= \{ V \in \mathcal{U}(\mathbb{C}^d) : V \mathcal{P} V^\dagger = \mathcal{P} \},\\
    \mathcal{C}_n &= \{ V \in \mathcal{U}((\mathbb{C}^d)^{\otimes n}) : V \mathcal{P}_n V^\dagger = \mathcal{P}_n \}.
\end{align*}
It is easy to verify that $
\mathcal{P}_n\subset\mathcal{C}_n\subset\mathcal{U}((\mathbb{C}^d)^{\otimes n}).$

A pure  state is called a stabilizer state if it is a common eigenstate $1$ of a maximal Abelian subgroup of the Clifford group $\mathcal{C}$. 
A mixed state is called a stabilizer state if it can be expressed as a convex mixture of pure stabilizer states. 
Otherwise, it is called a magic state or a nonstabilizer state. 
We denote by $\mathrm{STAB}$ the set of all stabilizer states \cite{Veitch2014}.

The Gottesman--Knill theorem states that any quantum circuit that starts in a stabilizer state, consists only of Clifford gates, and ends with Pauli measurements can be efficiently simulated on a classical computer \cite{Gottesman1997,Gottesman1999,Aaronson2004}.

An $n$-qudit Clifford circuit is a circuit composed of unitary operators from $\mathcal{C}_n$ and can be generated by the gate set $\{Z, G_{\rm P}, F, \text{CSUM}\}$, where the phase gate $ G_{\rm P}$, the discrete Fourier transform $F$, and the controlled-SUM gate are given by
\begin{align*}
   G_{\rm P} =& \sum_{j \in \mathbb{Z}_d} \tau^{j^2} |j\rangle\langle j|, \qquad
    F = \frac{1}{\sqrt{d}}\sum_{j,k \in \mathbb{Z}_d} \omega^{jk} |k\rangle\langle j|, \\
    \text{CSUM} =& \sum_{j \in \mathbb{Z}_d} |j\rangle\langle j| \otimes X^j=\sum_{j_1,j_2 \in \mathbb{Z}_d} \bigl| j_1 ,  j_1 +j_2 \bigr\rangle \langle j_1, j_2|.
\end{align*}
For qubits ($d=2$), CSUM reduces to the familiar CNOT gate, and for qutrits ($d=3$) it takes the form
\begin{equation}\label{CSUM3}
    \mathrm{CSUM}_3 
    = |0\rangle\!\langle 0| \otimes \mathbf{1}
    + |1\rangle\!\langle 1| \otimes X
    + |2\rangle\!\langle 2| \otimes X^{2}.
\end{equation}



\subsection{Discrete Wigner function and mana}

Let $\rho$ be a quantum state on a $d$-dimensional Hilbert space. Its discrete Wigner function at the phase-space point $(k,l) \in \mathbb{Z}_d \times \mathbb{Z}_d$ is defined as \cite{Wootters1987,Gibbons2004}
\begin{align*}
W_\rho(k,l) := \frac{1}{d}\,\mathrm{tr}(\rho A_{k,l}),
\end{align*}
where $A_{k,l} := D_{k,l} A_{0,0} D_{k,l}^\dagger$ is the phase-space point operator and $A_{0,0} := \frac{1}{d} \sum_{k,l \in \mathbb{Z}_d} D_{k,l}$ is the discrete parity operator.
The discrete Wigner function of the state $\rho$ can be expressed as a  $d\times d$ matrix $\boldsymbol{W}_{\rho}=\big(W_{\rho}({k,l})\big).$

The discrete Wigner function possesses the following fundamental properties \cite{Gibbons2004,Gross2006,Veitch2014,Wang2019}:

(1) Discrete Hudson's theorem.  
A pure state $|\psi\rangle$ is a stabilizer state if and only if its Wigner function is nonnegative everywhere:
\begin{equation}
   W_{|\psi\rangle}(k,l)\geq 0, \qquad \forall\, k,l \in \mathbb{Z}_d . \nonumber
\end{equation}

(2) The discrete Wigner function is a quasi-probability distribution in the sense that
\begin{equation}
   W_\rho(k,l)\in\mathbb{R}, 
   \qquad 
   \sum_{k,l\in\mathbb{Z}_d} W_\rho(k,l) = 1 , \nonumber
\end{equation}
but $W_\rho(k,l)$ may take negative values.

(3) The quantum state $\rho$ is uniquely determined by its discrete Wigner function in the sense that
\begin{equation}
   \rho = \sum_{k,l\in\mathbb{Z}_d} W_\rho(k,l)\, A_{k,l}, \nonumber
\end{equation}
where the phase-space point operators $\{A_{k,l}\}$ form a Hermitian, trace-one, orthogonal, and complete operator basis, satisfying
\begin{align}
&A_{k,l}^\dagger = A_{k,l}, \qquad \qquad \qquad \quad\mathrm{tr}(A_{k,l}) = 1, \label{Aklpro1}\\
&\mathrm{tr}(A_{k,l} A_{k',l'}) = d\,\delta_{k,k'}\delta_{l,l'}, \quad \sum_{k,l} A_{k,l} = d\,\mathbf{1}. \label{Aklpro2}
\end{align}

(4) If $\rho$ is displaced by a Heisenberg--Weyl operator $D_{m,n}$, then
\begin{equation}
   W_{D_{m,n}\rho D_{m,n}^\dagger}(k,l) 
   = W_\rho(k-m,\,l-n) \pmod d . \nonumber
\end{equation}

These properties make the negativity of the Wigner function a natural indicator of  magic (non-stabilizerness).  
A common quantifier is the sum negativity, defined as \cite{Veitch2014}
\begin{align}
\mathrm{sn}(\rho) \;:=\; \sum_{ W_\rho(k,l)<0} \bigl|W_\rho(k,l)\bigr| \;=\; \tfrac{1}{2}\Big(\sum_{k,l\in\mathbb{Z}_d} \bigl|W_\rho(k,l)\bigr| - 1\Big).  \nonumber
\end{align} 

While $\mathrm{sn}(\rho)$ directly measures the total weight of negative quasiprobabilities,  
it is often more convenient to work with its logarithmic form, known as the mana \cite{Veitch2014}
\begin{align}
\mathrm{Mana}(\rho):= \log\!\big(2\,\mathrm{sn}(\rho)+1\big)= \log\Big(\sum_{k,l\in\mathbb{Z}_d} \bigl|W_\rho(k,l)\bigr|\Big).  \nonumber
\end{align}

For a bipartite state $\rho_{ab}$ on the composite Hilbert space $\mathcal{H}_a \otimes \mathcal{H}_b$, where $\dim \mathcal{H}_a = d_a$ and $\dim \mathcal{H}_b = d_b$, the
mana can be similarly defined as
\begin{align*}
\mathrm{Mana}(\rho_{ab}) \;:= \log\bigg(\sum_{k,l,k^{\prime},l^{\prime}}\frac{1}{d_ad_b} \bigl|\mathrm{tr}(\rho_{ab} (A_{k,l}\otimes A_{k^{\prime},l^{\prime}}))\bigr|\bigg).
\end{align*}

The mana satisfies several useful properties \cite{Veitch2014}: 

(1) $\mathrm{Mana}(\rho) \ge 0$, and equality holds if $\rho$ is a stabilizer state.

(2) $\mathrm{Mana}(\rho)$  is convex in  $\rho$.

(3) The mana is invariant under Clifford unitaries in the sense that, for any Clifford unitary $V$,
    \begin{equation}\label{invariance}
        \mathrm{Mana}(V \rho V^\dagger) = \mathrm{Mana}(\rho).
    \end{equation}

(4) The mana is monotonic under stabilizer-preserving channels in the sense that, if $\Lambda$ is such a channel,
    \begin{equation}
        \mathrm{Mana}(\Lambda(\rho)) \le \mathrm{Mana}(\rho). \nonumber
    \end{equation}

(5) The mana is additive in the sense that, for any quantum states $\rho_1$ and $\rho_2$,
    \begin{equation}\label{additivity}
        \mathrm{Mana}(\rho_1 \otimes \rho_2) = \mathrm{Mana}(\rho_1) + \mathrm{Mana}(\rho_2).
    \end{equation}  

Beyond these general structural properties, it is also useful to establish 
explicit upper bounds on the mana of arbitrary quantum states. 
The following proposition provides such a bound  in terms of the purity of the state.

\begin{proposition}\label{P1}
For any quantum state $\rho$, 
\begin{align}
\mathrm{Mana}(\rho) \leq  \tfrac{1}{2}\log\!\bigl(d\, {\rm tr}(\rho^2)\bigr).  \label{Manarholeq}
\end{align}
\end{proposition}

For the proof, see Appendix \ref{proof_prop1}.

From Proposition~\ref{P1} and the convexity of the mana, it follows that the maximal mana is attained on pure states. In particular, for a pure state $\rho = |\psi\rangle\langle\psi|$, we have
\begin{align}
    \mathrm{Mana}(|\psi\rangle) \le \frac{1}{2} \log d. \label{Maxmana}
\end{align}
In the following, we investigate whether this bound in Eq. \eqref{Maxmana} can be saturated for maximally coherent states 
\begin{align}
    |\psi(\theta_1,\dots,\theta_{d-1})\rangle
= \frac{1}{\sqrt{d}} \sum_{j=0}^{d-1} e^{{\rm i} \theta_j} |j\rangle, \nonumber
\end{align}
where $\theta_0=0$ and $\theta_j\in[0,2\pi)$.

By direct calculation, for $d=3$, the maximal mana $\mathrm{Mana} = \tfrac{1}{2}\log 3$ is attained for 
$$(\theta_1,\theta_2) \in \{(2\pi/3,0), (0,2\pi/3), (4\pi/3,4\pi/3)\}.$$
For $d=5$, the maximal mana $\mathrm{Mana} = \tfrac{1}{2}\log 5$ occurs at
\begin{align*}
    (\theta_1,\theta_2,\theta_3,\theta_4) = (6\pi/5, 4\pi/5, 4\pi/5, 6\pi/5).
\end{align*}

These results demonstrate that the upper bound in Eq.~\eqref{Maxmana} is attainable for certain maximally coherent states in low dimensions. Whether this bound can be saturated for arbitrary odd prime $d$ remains an open question for further investigation.


\subsection{Discrete beamsplitters}
In this subsection, we review discrete  beamsplitters and their fundamental properties, and derive their action on discrete phase-space point operators as well as the corresponding expectation values.

For a bipartite system  $\mathcal{H}_a \otimes \mathcal{H}_b = \mathbb{C}^d \otimes \mathbb{C}^d$ with  prime local dimension $d$, a \emph{discrete lossless   beamsplitter} is defined by the unitary operator \cite{kbu2023,KBU22023}
\begin{equation}
    B_{t,r} = \sum_{j_1,j_2 \in \mathbb{Z}_d} \bigl| t j_1 + r j_2, \; t j_2 - r j_1 \bigr\rangle \langle j_1, j_2|,   \nonumber
\end{equation}
where $t,r \in \mathbb{Z}_d$ satisfy 
\begin{align}
    t^2 + r^2 = 1 \ (\mathrm{mod}\ d), \qquad t,r \neq 0 \ (\mathrm{mod}\ d), \nonumber
\end{align}
and $\{|j_1,j_2\rangle\}$ denotes the standard computational basis. Here $t^2$ and $r^2$ correspond to the transmission and reflection coefficients, respectively. For $d<7$, only trivial   beamsplitters exist, i.e., those with $t=0$ or $r=0$.

Motivated by the above construction, one can define a more general class of discrete   beamsplitters that encompasses all invertible linear transformations on the computational basis. 

For a bipartite system  $\mathcal{H}_a \otimes \mathcal{H}_b = \mathbb{C}^d \otimes \mathbb{C}^d$ with  prime local dimension $d$,
the \emph{generalized discrete   beamsplitter} is defined by the unitary operator
\begin{align}
    B_G = \sum_{j_1,j_2 \in \mathbb{Z}_d} \bigl| g(\delta j_1 - \gamma j_2), \; g(\alpha j_2 - \beta j_1) \bigr\rangle \langle j_1, j_2|,  \label{GDBS}
\end{align}
where $G$ is an invertible transformation matrix
\begin{align*}
    G = \begin{pmatrix} \alpha & \beta \\ \gamma & \delta \end{pmatrix}, \qquad
    \det G \not\equiv 0 \ (\mathrm{mod}\ d),
\end{align*}
and $g = (\det G)^{-1} \ (\mathrm{mod}\ d)$.

The  beamsplitter $B_G$ includes  the swap gate $S=\sum_{j_1,j_2\in\mathbb{Z}_d}|j_2,j_1\rangle\langle j_1,j_2|$ and   controlled-SUM gate $\mathrm{CSUM}$ as  special cases with
\begin{align}
   G_{\rm S} = \begin{pmatrix} 0 & 1 \\ 1 &0 \end{pmatrix}, \qquad G_{\mathrm{CSUM}} = \begin{pmatrix} 1 & -1+d \\ 0 &1 \end{pmatrix}.   \nonumber
\end{align}

In the two-qutrit system, there are four non-zero invertible parameter matrices $G$ as follows:
\begin{align*}
G_1 =&G_{\mathrm{CSUM}_3}= \begin{pmatrix} 1 & 2 \\ 0 &1 \end{pmatrix},\quad
    G_2 = \begin{pmatrix} 1 & 0 \\ 2 &1 \end{pmatrix},\\
    G_3 =& \begin{pmatrix} 0& 1 \\ 1 &2 \end{pmatrix},\quad\qquad\qquad\,\,\,\,
    G_4 = \begin{pmatrix} 2 & 1 \\ 1 &0 \end{pmatrix}.   
\end{align*}
Specifically, the qutrit  beamsplitters $B_{G_1}$ and $B_{G_2}$ are controlled-SUM gates, with the control qutrit on $\mathcal{H}_a$ and $\mathcal{H}_b$, respectively. The remaining two  $B_{G_3}$ and $B_{G_4}$  are composed of a swap gate and two types of controlled-SUM gates.

The  generalized discrete   beamsplitters \(B_G\) satisfy the following properties \cite{Feng2025}:

(1) If the input state \(\rho_{ab}^{\mathrm{in}}\) is incoherent in the computational basis of $\mathbb{C}^d \otimes \mathbb{C}^d$, then the output state 
\(\rho_{ab}^{\mathrm{out}} = B_G \, \rho_{ab}^{\mathrm{in}} \, B_G^\dagger\) is separable. However, the converse does not hold in general.  For example, 
 the maximally coherent product state $|+\rangle \otimes |+\rangle = \frac{1}{d} \sum_{j_1,j_2 \in \mathbb{Z}_d} |j_1,j_2\rangle$ is invariant under $B_G$,
\begin{align}
    B_G \bigl(|+\rangle \otimes |+\rangle\bigr) = |+\rangle \otimes |+\rangle. \nonumber
\end{align}

(2) \(B_G\) is a Clifford operator. In particular, for any discrete Heisenberg--Weyl operators \(D_{k_1,l_1}\) and \(D_{k_2,l_2}\),
\begin{align*}
    &B_G \left(D_{k_1,l_1} \otimes D_{k_2,l_2}\right) B_G^\dagger \nonumber\\
    = &D_{g(\delta k_1 - \gamma k_2), \, \alpha l_1 + \beta l_2} \otimes D_{g(\alpha k_2 - \beta k_1), \, \delta l_2 + \gamma l_1}.
\end{align*}
Moreover, for any discrete Heisenberg--Weyl operator \(D_{k,l}\),
\begin{align}
    B_G^\dagger \bigl(D_{k,l} \otimes \mathbf{1}\bigr) B_G = D_{\alpha k,\, g\delta l} \otimes D_{\beta k,\,-g\gamma l}.  \nonumber
\end{align}

(3) Suppose \(\beta \delta \neq 0\) (or \(\alpha \gamma \neq 0\)) and the input state takes the form 
\(\rho_{ab}^{\mathrm{in}} = \rho \otimes |j\rangle\langle j|\) (or \(|j\rangle\langle j| \otimes \rho\)).
Then the reduced states $\rho_a^{\mathrm{out}} = \mathrm{tr}_b ( B_G \rho_{ab}^{\mathrm{in}} B_G^\dagger)$,  $ 
    \rho_b^{\mathrm{out}} = \mathrm{tr}_a ( B_G \rho_{ab}^{\mathrm{in}} B_G^\dagger)$ of the output state  are both diagonal  in the standard computational basis.


These results characterize the action of $B_G$ on quantum states and the discrete Heisenberg--Weyl operators. We now turn to its action on the discrete phase-space point operators.

\begin{proposition}\label{prop2}
In a bipartite system  $\mathcal{H}_a \otimes \mathcal{H}_b = \mathbb{C}^d \otimes \mathbb{C}^d$ with odd prime $d$, the generalized discrete   beamsplitter $B_G$ defined by Eq. \eqref{GDBS} satisfies the following properties:

(1) For any phase-space point operators $A_{k_1,l_1},\, A_{k_2,l_2},$  
\begin{align*}
&B_G\bigl(A_{k_1,l_1}\otimes A_{k_2,l_2}\bigr)B_G^\dagger \nonumber\\
=& A_{\,g(\delta k_1-\gamma k_2),\;\alpha l_1+\beta l_2} \otimes A_{\,g(\alpha k_2-\beta k_1),\;\delta l_2+\gamma l_1}.  
\end{align*}

(2) For any phase-space point operators $A_{k,l},$
\begin{align}
&B_G^\dagger(A_{k,l}\otimes {\bf1})B_G \nonumber\\
=& \frac{1}{d} \sum_{m,n\in \mathbb{Z}_d} \omega^{\,lm - kn} (D_{\alpha m, g \delta n} \otimes D_{\beta m, -g \gamma n}), \label{BGAkl1} \\
&B_G^\dagger ({\bf1} \otimes A_{k,l}) B_G\nonumber\\
=& \frac{1}{d} \sum_{m,n \in \mathbb{Z}_d} \omega^{\,lm - kn} 
(D_{-\gamma m, g \beta n} \otimes D_{-\delta m, -g \alpha n}).  \label{BGAkl2}
\end{align}

\end{proposition}

For the proof, see Appendix \ref{proof_prop2}.

In the following proposition, we investigate the implications of the action of $B_G$ on expectation values when the input state is of the product form $\rho \otimes |0\rangle\langle 0|$.

\begin{proposition}\label{prop3}
For any phase-space point operators $A_{k,l}$ and any generalized discrete   beamsplitter $B_G$ with $\beta\delta\neq0$, 
\begin{align*}
\operatorname{tr}\Big((\rho \otimes |0\rangle\langle 0|) B_G^\dagger (A_{k,l} \otimes \mathbf{1}) B_G \Big) =& \rho_{j_0 j_0},\\
\operatorname{tr}\Big((\rho \otimes |0\rangle\langle 0|) B_G^\dagger ({\bf1} \otimes A_{k,l}) B_G \Big) =& \rho_{j_1 j_1},
\end{align*}
where $j_0 =k(g\delta)^{-1}= k{\delta}^{-1}\det G \,\ (\mathrm{mod}\ d) $ and $j_1= k\,(g\beta)^{-1}= k{\beta}^{-1}\det G \,\  (\mathrm{mod}\ d)$.
\end{proposition}

For the proof, see Appendix \ref{proof_prop3}.


\section{Mutual Mana as Magic Correlations}
In this section,  we introduce mutual mana as a measure of magic correlations and  present a systematic study of its fundamental properties.  We demonstrate that a generalized discrete   beamsplitter can completely transform local magic into inter-system magic correlations. We establish an upper bound on the mutual mana for maximally coherent states and show that this bound is tight in dimensions 3 and 5. Furthermore, we derive analytical expressions for the mutual mana of noisy real qutrit pure states and maximally coherent states after the action of the discrete   beamsplitter. We analyze the dependence of mutual mana on state parameters and identify the conditions under which it is maximized, supported by numerical examples and graphical illustrations.


\begin{definition}
For a bipartite quantum state $\rho_{ab}$ with reduced states $\rho_a = \mathrm{tr}_b(\rho_{ab})$ and $\rho_b = \mathrm{tr}_a(\rho_{ab})$, the \emph{mutual mana} of $\rho_{ab}$ is defined as
\begin{align}
    \mathcal{M}_{\rm mana}(\rho_{ab}) :=  \mathrm{Mana}(\rho_{ab}) -  \mathrm{Mana}(\rho_a) -  \mathrm{Mana}(\rho_b).  \nonumber
\end{align}
This quantity characterizes the correlations of magic resources between the two subsystems of $\rho_{ab}$.
\end{definition}


In the following, we summarize several basic properties of the mutual mana.

\begin{proposition}\label{prop_mutmana}
The mutual mana $\mathcal{M}_{\rm mana}(\rho_{ab})$ satisfies the following properties:

(1) If $\rho_{ab} = \rho_a \otimes \rho_b$ is a product state, then
\begin{align}
    \mathcal{M}_{\rm mana}(\rho_{ab}) = 0.  \nonumber
\end{align}

(2) For any Clifford operators $C_1$ acting on $\mathcal{H}_a$ and $C_2$ acting on $\mathcal{H}_b$,
\begin{align}
    \mathcal{M}_{\rm mana}\Big( (C_1 \otimes C_2) \rho_{ab} (C_1 \otimes C_2)^\dagger \Big) = \mathcal{M}_{\rm mana}(\rho_{ab}).   \nonumber
\end{align}
\end{proposition}

For the proof, see Appendix  \ref{proof_prop4}.

Notably, being a product state is not a necessary condition for the disappearance of mutual mana, as will be illustrated in Example~5. 


The following theorem establishes that a generalized discrete beamsplitter fully converts the local magic of an input state into magic correlations quantified by mutual mana shared between the two subsystems.

\begin{theorem}\label{theorem}
For any state $\rho$ on $\mathcal{H}_a$ and a generalized discrete   beamsplitter $B_G$ with $\beta\delta\neq0$ acting on $\mathcal{H}_a \otimes \mathcal{H}_b = \mathbb{C}^d \otimes \mathbb{C}^d$,
\begin{align}\label{prop4item3}
    \mathcal{M}_{\rm mana}\Bigl( B_G (\rho \otimes |0\rangle\langle 0|) B_G^\dagger \Bigr)=\mathrm{Mana}(\rho).  
\end{align}    
\end{theorem}

For the proof, see Appendix \ref{proof_thm}.

In this process, all of the initial local magic resource  of $\rho$ is transferred into mutual mana, while the output local magic of each subsystem remains zero. 
This result provides an operational interpretation of mutual mana as the correlation form of quantum magic generated by the beamsplitter.

 Next, we turn to a fundamental family of quantum states: the maximally coherent states. These states play vital roles in quantum cryptographic protocols and admit optimal cloning procedures \cite{Fan2003,Scarani2005}.

\begin{proposition}
Consider the maximally coherent state
 \begin{equation}
  |\psi_{\boldsymbol{\theta}}\rangle= \frac{1}{\sqrt{d}} \sum_{j \in \mathbb{Z}d} e^{{\rm i}\theta_j} |j\rangle \in \mathbb{C}^d, \nonumber
\end{equation}
where $\theta_0=0$ and   $\boldsymbol{\theta} = (\theta_1, \dots, \theta_{d-1}) \in \mathbb{R}^{d-1}$.
 Let the input state be $|\psi_{\boldsymbol{\theta}}\rangle \otimes |0\rangle$, and let it be acted upon by a discrete beamsplitter $B_G$ with parameters satisfying $\beta\delta \neq 0$. Then
 \begin{equation}
 \max_{\boldsymbol{\theta} \in \mathbb{R}^{d-1}} \; \mathcal{M}_{\rm mana}\Big(B_G(|\psi_{\boldsymbol{\theta}}\rangle \otimes |0\rangle)\Big) \leq \frac{1}{2}\log(d).   \nonumber
  \end{equation}
 Moreover, the equality is attained in dimensions $d=3$ and $d=5$.
 \end{proposition}

For the proof, see Appendix \ref{proof_prop5}.

We examine the following qutrit states as representative examples to illustrate Theorem  1.

Consider the noisy input qutrit state $\rho_{\phi,p}=p|\phi\rangle\langle\phi|+ (1-p) {\bf 1}/3$, where $|\phi\rangle = \sum^2_{i=0}\mu_i|i\rangle$ is an arbitrary qutrit state ($\mu_i \in \mathbb{C}$, $\sum_i |\mu_i|^2=1$).  The mutual mana of  the output state $\rho_{\phi,p}^{\rm out}=\mathrm{CSUM}_3(\rho_{\phi,p}\otimes|0\rangle\langle 0|)\mathrm{CSUM}^{\dag}_3$ is 
\begin{align}
   \mathcal{M}_{\rm mana}(\rho_{\phi,p}^{\rm out})=\mathrm{Mana}(\rho_{\phi,p})=\log\Big(\sum_{k,l} \bigl|W_{\rho_{\phi,p}}(k,l)\bigr|\Big).\nonumber
\end{align}
The Wigner function matrix of $\rho_{\phi,p}$ is
\begin{align}
   \boldsymbol{W}_{\rho_{\phi,p}} =p \boldsymbol{W}_{|\phi\rangle}+\frac{1-p}{9}\boldsymbol{J}_3, \nonumber
\end{align}
    where $\boldsymbol{J}_3$ denotes the $3\times 3$ all-ones matrix,
 and   $\boldsymbol{W}_{|\phi\rangle}$ is  the  Wigner function matrix of  the qutrit state  $|\phi\rangle$:
{\footnotesize\begin{align*}
   \frac{1}{3} \begin{pmatrix} |\mu_0|^2+2{\rm Re}(\mu_1\mu^{*}_2), & |\mu_0|^2+2{\rm Re}(\mu_1\mu^{*}_2\omega),& |\mu_0|^2+2{\rm Re}(\mu_1\mu^{*}_2\omega^2) \\  |\mu_1|^2+2{\rm Re}(\mu_2\mu^{*}_0), & |\mu_1|^2+2{\rm Re}(\mu_2\mu^{*}_0\omega),& |\mu_1|^2+2{\rm Re}(\mu_2\mu^{*}_0\omega^2)\\|\mu_2|^2+2{\rm Re}(\mu_0\mu^{*}_1), & |\mu_2|^2+2{\rm Re}(\mu_0\mu^{*}_1\omega),& |\mu_2|^2+2{\rm Re}(\mu_0\mu^{*}_1\omega^2) \end{pmatrix}, 
\end{align*}}
with  $\omega=e^{{\rm i}2\pi/3}$ and ${\rm Re}(z)$  being the real part of  complex number $z$.  
This result can be obtained from  Theorem 1  and  direct calculation.



Specifically, when the input qutrit states have real coefficients, i.e., $\mu_i
\in\mathbb{R}$,  we obtain the following analytical expression.

\textbf{Example 1.}
Consider the noisy input qutrit state of the form
\begin{align*}
    \rho_{\phi,p} = p\,|\phi\rangle\langle\phi| + (1-p)\,\frac{\mathbf{1}}{3},
\end{align*}
where $|\phi\rangle = \sum^2_{i=0}\mu_i|i\rangle$ is a real pure state  with $\mu_i
\in\mathbb{R}$ and $\sum^2_{i=0}\mu_i^2=1$.
By direct calculation, the mutual mana of  the output state $\rho_{\phi,p}^{\rm out} = \mathrm{CSUM}_3 (\rho_{\phi,p} \otimes |0\rangle\langle 0|)\, \mathrm{CSUM}_3^\dagger$
is given by
{\footnotesize
\begin{align}
&\mathcal{M}_{\rm mana}(\rho^{\rm out}_{\phi,p})\nonumber\\
=&\log \Biggl[ \frac{1}{9} \Bigl( \Bigl| 2(1-p) + 6p( \mu_0^2 -  \mu_1 \mu_2 ) \Bigr|+
 \Bigl| 2 (1- p) + 6p(\mu_1^2 - \mu_0 \mu_2) \Bigr|  \nonumber\\
&\quad + \Bigl| 2(1-p) + 6p( \mu_2^2 - \mu_0 \mu_1 ) \Bigr| + \Bigl| 1 - p + 3p(\mu_1^2 + 2 \mu_0 \mu_2) \Bigr| \nonumber\\
&\quad + \Bigl| 1-p + 3p( \mu_0^2 + 2 \mu_1 \mu_2 ) \Bigr|  + \Bigl| 1 -p+ 3p(\mu_2^2 + 2 \mu_0 \mu_1 ) \Bigr| 
\Bigr) \Biggr].   \nonumber
\end{align}
}

\textbf{Example 2.}  
Consider the noisy input qutrit state
\begin{align*}
    \rho_{\psi(\theta_1,\theta_2),p} = p\,|\psi(\theta_1,\theta_2)\rangle\langle\psi(\theta_1,\theta_2)| + (1-p)\,\frac{\mathbf{1}}{3},
\end{align*}
where 
$|\psi(\theta_1,\theta_2)\rangle =\bigl(|0\rangle + e^{{\rm i}\theta_1}|1\rangle + e^{{\rm i}\theta_2}|2\rangle\bigr)/\sqrt{3}$
is a maximally coherent state with $\theta_i \in [0,2\pi)$.  
By direct calculation, the mutual mana of the output state 
$\rho_{\psi(\theta_1,\theta_2),p}^{\rm out} = \mathrm{CSUM}_3 (\rho_{\psi(\theta_1,\theta_2),p} \otimes |0\rangle\langle 0|)\, \mathrm{CSUM}_3^\dagger$
is given by
{\small
\begin{align*}
   &\mathcal{M}_{\rm mana}(\rho_{\psi(\theta_1,\theta_2),p}^{\rm out})\\
   =&\log\Bigg[\frac{1}{9}\Big(\big|1 + 2p\cos\theta_1\big| +\big|1 + 2p\cos\theta_2\big| \nonumber\\
   &+ \big|1 - p\cos\theta_1 + \sqrt{3}t\sin\theta_1\big| +\big|1 - p\cos\theta_2 + \sqrt{3}p\sin\theta_2\big| \nonumber\\
   &+ \big|1 - p\cos\theta_1 - \sqrt{3}p\sin\theta_1\big|  +\big|1 - p\cos\theta_2 - \sqrt{3}p\sin\theta_2\big|\Big) \nonumber\\
  &+ \big|1 + 2p\cos(\theta_1 - \theta_2)\big| \nonumber\\
  &+\big|1 - p\cos(\theta_1 - \theta_2) + \sqrt{3}p\sin(\theta_1 - \theta_2)\big| \nonumber\\
  &+\big|1 - p\cos(\theta_1 - \theta_2) - \sqrt{3}p\sin(\theta_1 - \theta_2)\big| \Big) \Bigg].
\end{align*}
}

\textbf{Example 3.} Consider  the noisy input qutrit state of the form
\begin{align}
    \rho_{\Phi_\lambda,p}=p|\Phi_\lambda\rangle\langle\Phi_\lambda|+ (1-p) \frac{\bf 1}{3},  \nonumber
\end{align}
where $p\in[0,1]$ and $|\Phi_{\lambda}\rangle = \lambda |0\rangle + \lambda |1\rangle + \sqrt{1-2\lambda^{2}}\,|2\rangle$ with $\lambda \in [0,1/\sqrt{2}]$.
By direct calculation, the mutual mana of the output state
$\rho_{\Phi_\lambda,p}^{\rm out} = \mathrm{CSUM}_3 \big(\rho_{\Phi_\lambda,p}\otimes|0\rangle\langle0|\big)\mathrm{CSUM}_3^{\dagger}$
is given by
{\small
\begin{align*}
&\mathcal{M}_{\rm mana}(\rho_{\Phi_\lambda,p}^{\rm out}) \nonumber \\
= &\log\Big[\frac{1}{9}\Big(3+6p\lambda(\lambda+2\sqrt{1-2\lambda^2}) \nonumber\\
&\qquad +2|1+p(2-9\lambda^2)| +4|1-p+3p\lambda(\lambda-\sqrt{1-2\lambda^2})|\Big)\Big].
\end{align*}
}

\begin{figure}[!t] 
\centering \includegraphics[width=0.5\textwidth]{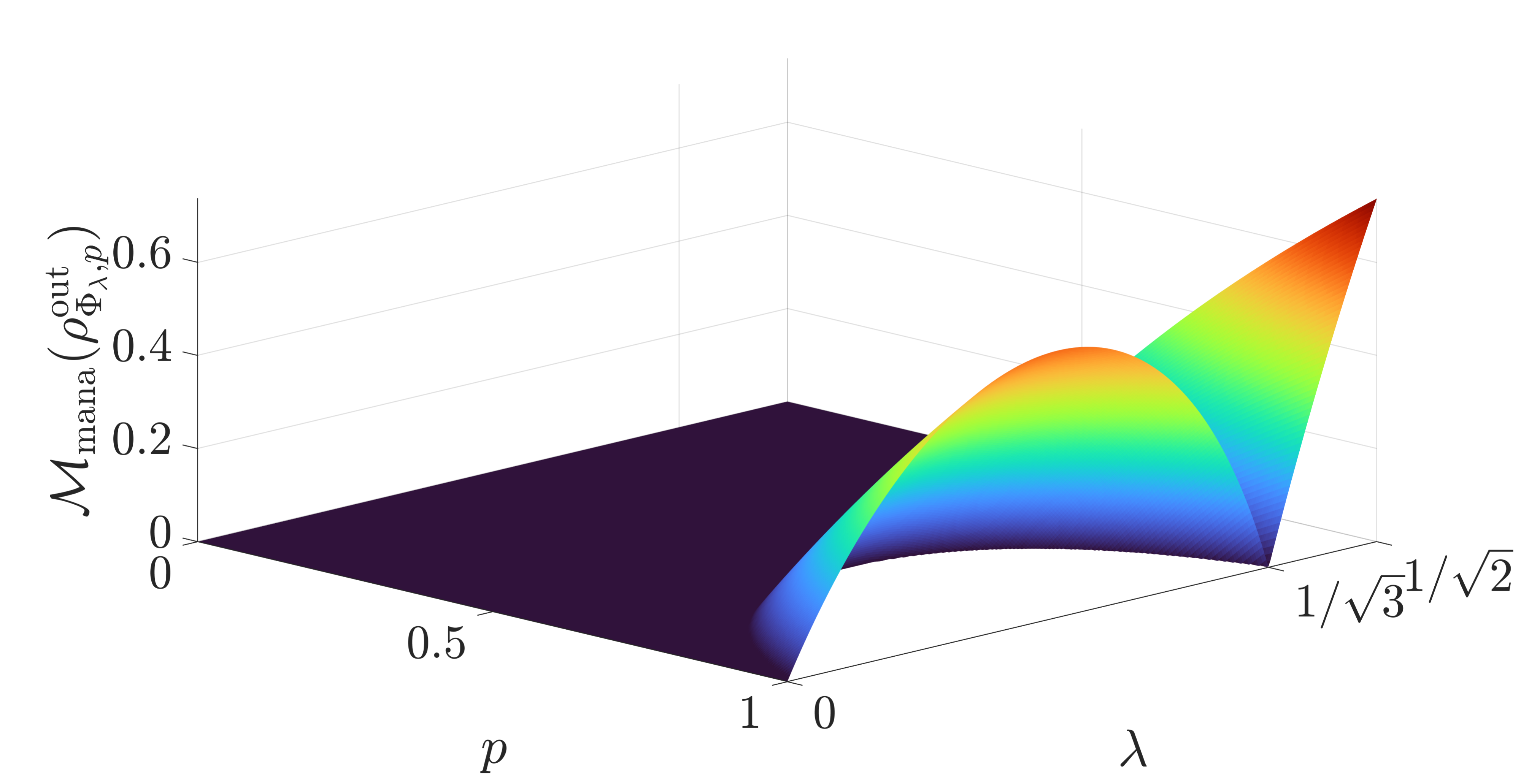} 
\caption{Landscape of mutual mana $\mathcal{M}_{\rm mana}(\rho_{\Phi_\lambda,p}^{\rm out})$ for the qutrit state 
    $\rho_{\Phi_\lambda,p}^{\rm out} = \mathrm{CSUM}_3\big(\rho_{\Phi_\lambda,p}\otimes|0\rangle\langle0|\big)
    \mathrm{CSUM}_3^{\dagger}$ as a function of $p$ and $\lambda$, 
    where $ \rho_{\Phi_\lambda,p}=p|\Phi_\lambda\rangle\langle\Phi_\lambda|+ (1-p) {\bf 1}/3$ and $|\Phi_{\lambda}\rangle = \lambda |0\rangle + \lambda |1\rangle + \sqrt{1-2\lambda^{2}}\,|2\rangle$.}\label{Figure_1} 
\end{figure}

Fig.~\ref{Figure_1} shows the variation of  mutual mana $\mathcal{M}_{\rm mana}(\rho_{\Phi_\lambda,p}^{\rm out})$ with parameters $p$ and $\lambda$, where it attains its maximum value of $\log(5/3)$ at  $p = 1$ and $\lambda = 1/\sqrt{2}$.

\textbf{Example 4.}
Consider the noisy input qutrit state of the form
\begin{align}
    \rho_{\psi_{\theta,p}}=p|\psi_\theta\rangle\langle\psi_\theta|+(1-p)\frac{{\bf1}}{3},  \nonumber
\end{align}
where $p\in[0,1]$ and $|\psi_\theta\rangle = \cos\theta\,|0\rangle + \sin\theta\,|1\rangle$ with $\theta\in[0,\frac{\pi}{2}]$.
By direct calculation, the mutual mana of the output state $\rho_{\psi_{\theta,p}}^{\rm out} 
= \mathrm{CSUM}_3 \big(\rho_{\psi_{\theta,p}}\otimes|0\rangle\langle0|\big)
\mathrm{CSUM}_3^{\dagger}$ is given by
\begin{align*}
&\mathcal{M}_{\rm mana}(\rho_{\psi_{\theta,p}}^{\rm out}) \nonumber\\
=& \log\!\left[\frac{1}{9}\Big(7 + 2p + \big| -2 + 2p + 3p\sin(2\theta) \big| + 3p\sin(2\theta)\Big)\right]\\
=&
\begin{cases}
\displaystyle
0,
& \text{if } p \leq \dfrac{2}{2 + 3\sin(2\theta)}, \\[10pt]
\displaystyle
\log\!\left[\frac{1}{9}\Big(5 + 4p + 6p\sin(2\theta)\Big)\right],
& \text{if } p > \dfrac{2}{2 + 3\sin(2\theta)}.
\end{cases}
\end{align*}

\begin{figure}[!t] 
\centering \includegraphics[width=0.5\textwidth]{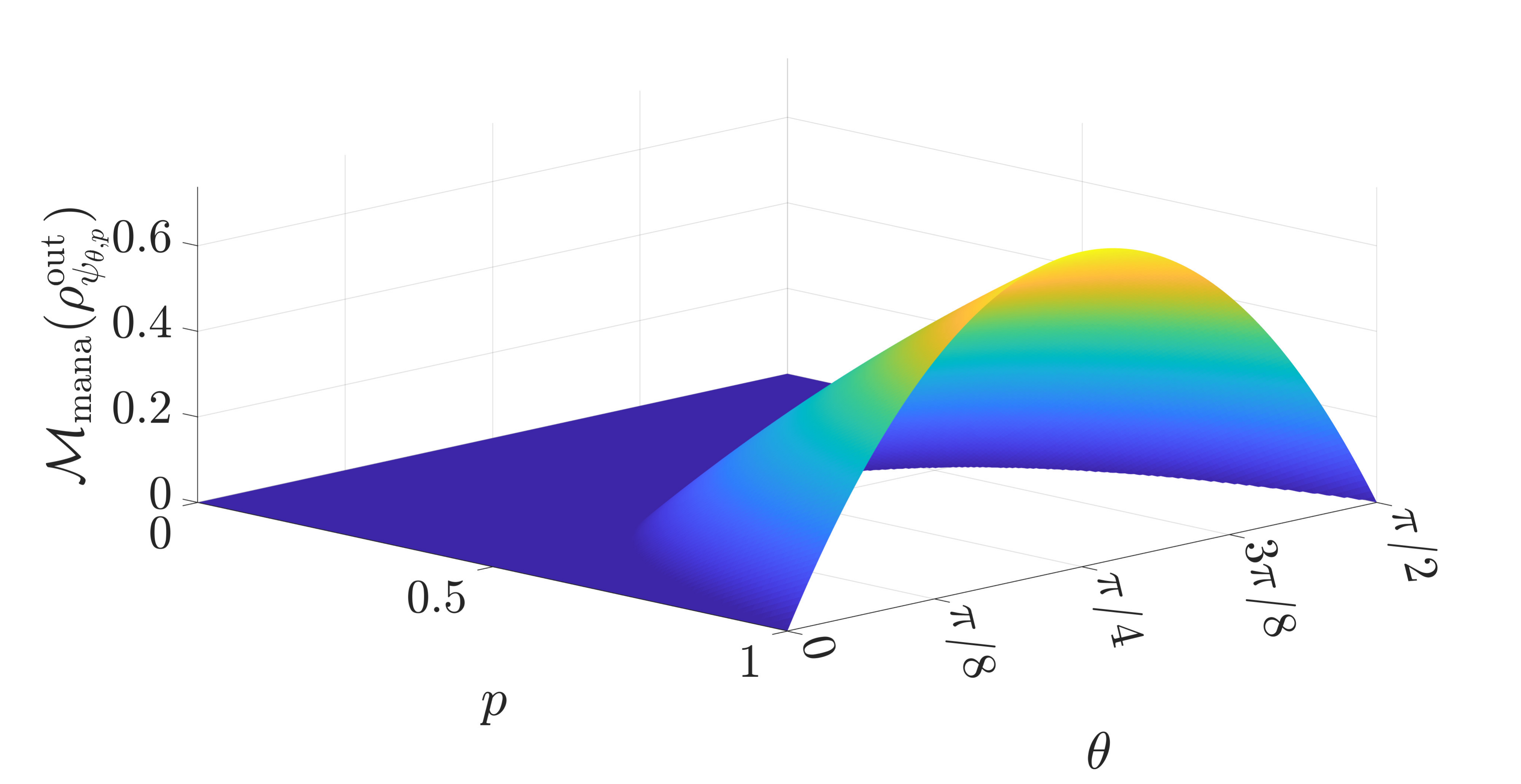} 
\caption{Landscape of mutual mana $\mathcal{M}_{\rm mana}(\rho_{\psi_{\theta,p}}^{\rm out})$ for the qutrit state 
    $\rho_{\psi_{\theta,p}}^{\rm out} = \mathrm{CSUM}_3\big(\rho_{\psi_{\theta,p}}\otimes|0\rangle\langle0|\big)
    \mathrm{CSUM}_3^{\dagger}$ as a function of $p$ and $\theta$, 
    where $\rho_{\psi_{\theta,p}}=p|\psi_\theta\rangle\langle\psi_\theta|+(1-p){\bf1}/3$ and $|\psi_\theta\rangle = \cos\theta\,|0\rangle + \sin\theta\,|1\rangle$.}\label{Figure_2} 
\end{figure}

 Fig.~\ref{Figure_2} shows the variation  of the mutual mana    $\mathcal{M}_{\rm mana}(\rho_{\psi_{\theta,p}}^{\rm out})$  with  parameters $p$ and $\theta$.  $\mathcal{M}_{\rm mana}(\rho_{\psi_{\theta,p}}^{\rm out})$  vanishes when   $p \leq \dfrac{2}{2 + 3\sin(2\theta)}$, and attains its maximum $\log(5/3)$ at $p = 1$ and $\theta=\pi/4$.

\section{Comparison}
In this section, we review existing measures of correlations  that structurally parallel quantum mutual information and compare them with the mutual mana.

Recal  that the quantum mutual information  of  a bipartite quantum state $\rho_{ab}$ is defined as \cite{Nielsen2010}
\begin{align}
    I(\rho_{ab}) = S(\rho_{a}) + S(\rho_{b}) - S(\rho_{ab}),   \nonumber
\end{align}
where $S(\rho) = -\,{\rm tr}(\rho \log \rho)$ denotes the von Neumann entropy, and $\rho_{a} = {\rm tr}_{b}(\rho_{ab})$, $\rho_{b} = {\rm tr}_{a}(\rho_{ab})$ are the reduced states of $\rho_{ab}$  on subsystems $\mathcal{H}_a$ and $\mathcal{H}_b$, respectively.

Quantum mutual information  is a measure of  total correlations and serves as a fundamental building block for quantum discord \cite{Ollivier2001,Henderson2001,SLUO2008},  squashed entanglement \cite{Alicki2004,Brandao2011}, coherent information \cite{Schumacher1996,Lloyd1997}, and channel capacities \cite{Holevo2012}.


Inspired by quantum mutual information, magic correlations including mutual stabilizer $\alpha$-Rényi entropy and mutual $L^1$-norm magic have attracted widespread interest due to their straightforward computation, lack of complex optimization, and well-defined formulation in any finite-dimensional Hilbert space.

Recall that  the mutual stabilizer  $\alpha$-Rényi    entropy
for a bipartite quantum state $\rho_{ab}$  is defined  as 
\cite{White2021,Tarabunga2023,Lopez2024,Qian2025,Tarabunga2025,Frau2025}
\begin{align}
    \mathcal{M}_{\mathrm{SRE}_\alpha}
(\rho_{ab})
    = M_{\mathrm{SRE}_\alpha}(\rho_{ab})
    - M_{\mathrm{SRE}_\alpha}(\rho_{a})
    - M_{\mathrm{SRE}_\alpha}(\rho_{b}),   \nonumber
\end{align}
where $M_{\mathrm{SRE}_\alpha}(\rho)$ denotes the  stabilizer  $\alpha$-Rényi entropy of a quantum state $\rho$.  
For an $n$-qudit state $\rho$, it is given by \cite{Leone2022,Haug2023}
{\small
\begin{align}\label{MSRE}
    M_{\mathrm{SRE}_\alpha}(\rho)=\frac{\alpha}{1-\alpha}\log\left(\sum_{P\in\mathcal{P}_n}\left(\frac{|{\rm tr}(\rho P)|^2}{d^n{\rm tr}\rho^2}\right)^\alpha\right)^{\frac{1}{\alpha}}-n\log{d},  
\end{align}
}
where $\mathcal{P}_n$ is the $n$-qudit discrete Heisenberg-Weyl group  defined by  Eq. (\ref{gpg}).

 The magic quantifier $ M_{\mathrm{SRE}_\alpha}(\rho)$ defined by Eq. ({\ref{MSRE}})  possesses several important   properties such as Clifford invariance and additivity. For any pure states $|\phi\rangle$ in $\mathbb{C}^d$, $ M_{\mathrm{SRE}_\alpha}(|\phi\rangle)$ reaches its minimum value of 0 if and only if $|\phi\rangle$ is a stabilizer state. 
We remark that $ M_{\mathrm{SRE}_\alpha}(\rho)$  is not convex.

The mutual $L^1$-norm magic $ \mathcal{M}_{L^1}(\rho_{ab}) $ serves as an alternative quantifier of magic correlations constructed from the $L^1$-norm of characteristic function and defined as \cite{Feng2025}
\begin{align}
    \mathcal{M}_{L^1}(\rho_{ab})
    = \log M_{L^1}(\rho_{ab})
    - \log M_{L^1}(\rho_{a})
    - \log M_{L^1}(\rho_{b}),   \nonumber
\end{align}
where $M_{L^1}(\rho_{x})$ ($x=a,\,b$) and $M_{L^1}(\rho_{ab})$ denote the $L^1$-norm magic of the reduced and global states, respectively. They are given by \cite{Dai2022}
 \begin{align}
   M_{L^1}(\rho_{x})
    &= \sum_{k,l\in\mathbb{Z}_d}
       \big|{\rm tr}\!\left(\rho_{x} D_{k,l}\right)\big|, 
       \qquad (x=a,\, b),  \label{ML1rho} \\
    M_{L^1}(\rho_{ab})
   & = \sum_{k_1,k_2,l_1,l_2\in\mathbb{Z}_d}
       \big|{\rm tr}\!\left(\rho_{ab}\,
       D_{k_1,l_1}\!\otimes\! D_{k_2,l_2}\right)\big|.\notag
\end{align}

 The magic quantifier $M_{L^1}(\rho)$ defined by Eq. ({\ref{ML1rho}}) has some nice properties such as Clifford invariance and convexity. It is uniquely minimized among all states  $\rho$ in $\mathbb{C}^d$ by the maximally mixed state $\mathbf{1}/d$, while for pure states $|\phi\rangle$ in $\mathbb{C}^d$, $M_{L^1}(|\phi\rangle)$ reaches its minimum value if and only if $|\phi\rangle$ is a stabilizer state. Consequently, the condition $M_{L^1}(\rho) >d$  provides an effective criterion for magic state detection.   
 

\begin{table*}[htbp]
  \centering
  \caption{Comparison of different  correlations generated by the qutrit  beamsplitter $\mathrm{CSUM}_3$.
  The   output state   $\rho_{\phi,p}^{\rm out}=\mathrm{CSUM}_3(\rho_{\phi,p}\otimes|0\rangle\langle 0|)\mathrm{CSUM}^{\dag}_3$    results from the noisy input state $\rho_{\phi,p}=p|\phi\rangle\langle\phi|+ (1-p) {\bf 1}/3$. Here, $h(p_1,p_2,p_3)=-\sum^3_{i=1} p_i\log p_i$ denotes the  Shannon entropy.}
  \label{tab1}
  \renewcommand{\arraystretch}{1.7}
  \begin{tabular}{  |c| c c c c| } 
    \hline
   &   $\rho_{S,p}^{\rm out}$ & $\rho_{N,p}^{\rm out}$ &  $\rho_{T,p}^{\rm out}$ &  $\rho_{H,p}^{\rm out}$\\
   \hline
   $I$& $\makecell{2h\big(\frac{1-p}{3},\frac{2+p}{6},\frac{2+p}{6}\big)\\-h\big(\frac{1-p}{3},\frac{1-p}{3},\frac{1+2p}{3}\big)}$
   &$\makecell{2h\big(\frac{2-p}{6},\frac{2-p}{6},\frac{1+p}{3}\big)\\-h\big(\frac{1-p}{3},\frac{1-p}{3},\frac{1+2p}{3}\big)}$ 
   & $2\log3-h\big(\frac{1-p}{3},\frac{1-p}{3},\frac{1+2p}{3}\big)$ &$\makecell{2h\big(\frac{2+p(1+\sqrt{3})}{6},\frac{4-p(1+\sqrt{3})}{12},\frac{4-p(1+\sqrt{3})}{12}\big)\\-h\big(\frac{1-p}{3},\frac{1-p}{3},\frac{1+2p}{3}\big)} $\\
   \hline
    $ \mathcal{M}_{\rm mana}$ &  $\max\big\{0,\log\big(\frac{7+8p}{9}\big)\big\}$ & $\max\big\{0,\log\big(\frac{5+10p}{9}\big)\big\}$ & $\max\big\{0,\log\big(\frac{1}{3}\big(1+4p\cos{\frac{\pi}{9}}\big)\big)\big\}$ & $\max\big\{0,\log\big(\frac{1}{9}\big(1+2p(1+3\sqrt{3})\big)\big)\big\}$\\ 
    \hline
     $\mathcal{M}_{L^1}$& $ \log\frac{3 + 12p}{(1 + p)^2}$ &$ \log\frac{3 + 12p}{(1 + p)^2}
$&$\log(3+6\sqrt{3}p)$&Eq. (\ref{ML1H})\\
    \hline
    $\mathcal{M}_{\mathrm{SRE}_2}$& $\log\frac{2+4p^2}{2+p^4}$&$\log\frac{2+4p^2}{2+p^4}$&$\log\frac{3+6p^2}{3+2p^4}$& Eq. (\ref{MSREH})\\
     \hline
  \end{tabular}
\end{table*} 

In order to gain a more concrete and intuitive  insight of mutual mana and its  relations with other magic correlations, we compare the amounts of these correlations generated by the qutrit beamsplitter for several representative qutrit states.

\textbf{Example 5.}
Consider the input qutrit pure states of the form
$$|\Phi_{\lambda}\rangle = \lambda |0\rangle + \lambda |1\rangle + \sqrt{1-2\lambda^{2}}\,|2\rangle,\qquad \lambda \in [0,1/\sqrt{2}].$$
After the action of the qutrit  beamsplitter $\mathrm{CSUM}_3$, the output states are given by
\begin{align*}
|\Phi_{\lambda}^{\rm out}\rangle
=&\mathrm{CSUM}_3(|\Phi_{\lambda}\rangle\otimes|0\rangle) \\
=&\lambda |00\rangle + \lambda |11\rangle + \sqrt{1-2\lambda^{2}}|22\rangle.
\end{align*}
The reduced states of the two subsystems are identical and given by
\begin{align*}
{\rm tr}_x(|\Phi_{\lambda}^{\rm out}\rangle\langle \Phi_{\lambda}^{\rm out}|)
=\lambda^2\sum^1_{i=0}|i\rangle\langle i|+(1-2\lambda^2)|2\rangle\langle2|,\,\, x=a, b.
\end{align*}
By direct calcution, we have
{\footnotesize\begin{align*}
I(|\Phi_{\lambda}^{\rm out}\rangle)=&-2\left(4\lambda^2\log\lambda+(1-2\lambda^2)\log(1-2\lambda^2)\right)
,\\
\mathcal{M}_{\mathrm{SRE}_2}(|\Phi_{\lambda}^{\rm out}\rangle)
=&\log\!\left(\frac{1 + f_1(\lambda)^2 + 2f_2(\lambda)^2+ f_3(\lambda)^2+ 4f_4(\lambda)^2}{1 + f_1(\lambda)^4 + 2f_2(\lambda)^4+ f_3(\lambda)^4 + 4f_4(\lambda)^4}\right),\\
\mathcal{M}_{L^1}\!\left(|\Phi_{\lambda}^{\mathrm{out}}\rangle\right) 
=& -2\log\Big(1 + f_1(\lambda) + f_3(\lambda) \Big) \nonumber\\
& + \log\Big[3\Big(1 + f_1(\lambda) + 2f_2(\lambda)+ f_3(\lambda)+ 4f_4(\lambda)\Big)\Big],\\
 \mathcal{M}_{\rm mana}(|\Phi_{\lambda}^{\rm out}\rangle)
=&\log\!\left(
\frac{1}{3}\!\left( 1 + 2f_1(\lambda) + 2f_2(\lambda )   + 4f_5(\lambda)\right)\right),
\end{align*}}
where $f_1(\lambda)=\big|1-3\lambda^2\big|$, $f_2(\lambda)=\lambda(\lambda + 2\sqrt{1 - 2\lambda^{2}})$, $f_3(\lambda)=\big|e^{\pi {\rm i}/3} - (1 + e^{\pi {\rm i}/3})^{2}\lambda^{2}\big|$, $f_4(\lambda)=\lambda\big|e^{2\pi {\rm i}/3}\lambda - (-1 + e^{\pi {\rm i}/3})\sqrt{1 - 2\lambda^{2}}\big|$, and $f_5(\lambda)=\lambda\big|\lambda - \sqrt{1 - 2\lambda^{2}}\big|$.

\begin{figure}[!t]
    \centering
    \begin{subfigure}{0.235\textwidth}
        \centering
        \includegraphics[width=\textwidth]{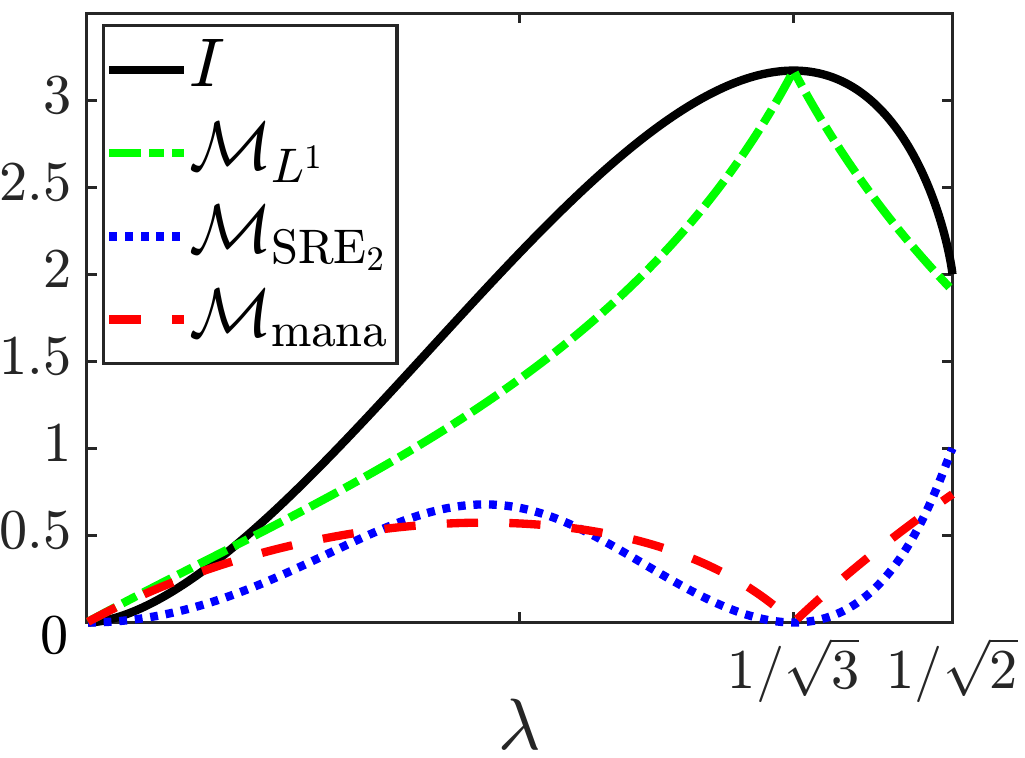}
        \caption{ $|\Phi_\lambda^{\mathrm{out}}\rangle$}
        \label{fig:3a}
    \end{subfigure}
    \hfill
    \begin{subfigure}{0.235\textwidth}
        \centering
        \includegraphics[width=\textwidth]{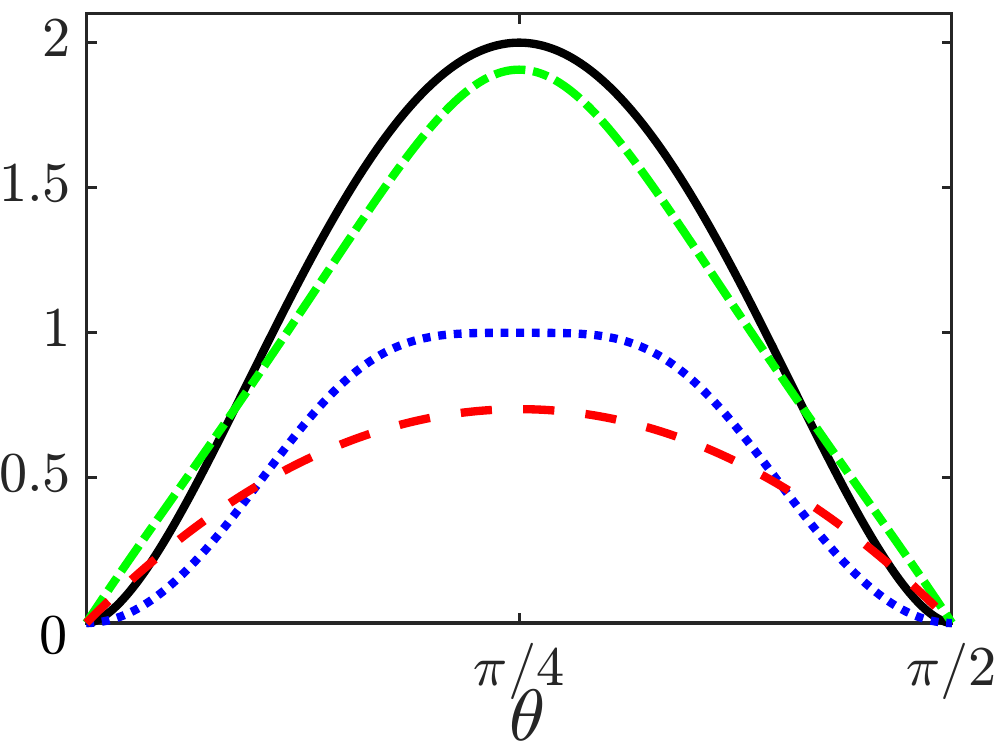}
        \caption{$|\psi_\theta^{\mathrm{out}}\rangle$}
        \label{fig:3b}
    \end{subfigure}
    \caption{Comparison of  quantum mutual information $I$ (black solid line), mutual $L^1$-norm magic  $\mathcal{M}_{L^1}$ (green  dash-dotted line), mutual stabilizer 2-Rényi entropy $\mathcal{M}_{\mathrm{SRE}_2}$ (blue dotted line),  and mutual mana $\mathcal{M}_{\rm mana}$ (red dashed line)  for the  output states: (a) $|\Phi_\lambda^{\mathrm{out}}\rangle =\mathrm{CSUM}_3(|\Phi_{\lambda}\rangle\otimes|0\rangle)= \lambda|00\rangle + \lambda|11\rangle + \sqrt{1-2\lambda^2}|22\rangle$ versus the parameter~$\lambda\in[0,1/\sqrt{2}]$  (b) $|\psi_\theta^{\mathrm{out}}\rangle
   = \mathrm{CSUM}_3(|\psi_{\theta}\rangle\otimes|0\rangle)=\cos\theta\,|00\rangle + \sin\theta\,|11\rangle$ versus the parameter~$\theta\in[0,\pi/2]$.
    }
    \label{fig:3}
\end{figure}

The variation of the four measures with the parameter~$\lambda$  is shown in Fig.~\ref{fig:3}(a).
When $\lambda=0$,  the input state $|\Phi_{0}\rangle=|2\rangle$ is an incoherent stabilizer  state, and the  output state $|\Phi^{\rm out}_{0}\rangle=|22\rangle$ is a product stabilizer state; consequently, all four  quantifiers of correlations  vanish.   When $\lambda=1/\sqrt{3}$,  the input state $|\Phi_{1/\sqrt{3}}\rangle=(|0\rangle+|1\rangle+|2\rangle)/\sqrt{3}$ is a maximally coherent  state, and the corresponding output state $|\Phi^{\rm out}_{1/\sqrt{3}}\rangle$  of  beamsplitter CSUM$_3$  achieves the maximum quantum mutual information and mutual  $L^1$-norm magic.  When $\lambda=1/\sqrt{2}$,  the input state $|\Phi_{1/\sqrt{2}}\rangle=(|0\rangle+|1\rangle)/\sqrt{2}$ possesses the largest  mana among all  $|\Phi_{\lambda}\rangle$, and the corresponding output state $|\Phi^{\rm out}_{1/\sqrt{2}}\rangle$  of  beamsplitter CSUM$_3$ achieves the maximum  mutual  mana and mutual stabilizer 2-Rényi entropy.

It is noteworthy that when $\lambda = 1/\sqrt{3}$, the output state $|\Phi^{\rm out}_{1/\sqrt{3}}\rangle$ exhibits vanishing mutual mana despite being entangled. This demonstrates that being a product state is  not a necessary condition for   the vanishing of  mutual mana.  
 


\textbf{Example 6.}
Consider the input qutrit pure states of the form
$$|\psi_\theta\rangle = \cos\theta\,|0\rangle + \sin\theta\,|1\rangle\in\mathbb{C}^3,\qquad \theta\in[0,\pi/2].$$ 
After the action of the qutrit  beamsplitter $\mathrm{CSUM}_3$, the output states are given by
\begin{align*}
    |\psi^{\rm out}_\theta\rangle
    =\mathrm{CSUM}_3(|\psi_\theta\rangle\otimes|0\rangle)
    =\cos\theta\,|00\rangle + \sin\theta\,|11\rangle.
\end{align*}
The reduced states of the two subsystems are identical and given by
\begin{align*}
   {\rm tr}_x(|\psi_\theta^{\rm out}\rangle\langle \psi_\theta^{\rm out}|)=\cos^2{\theta}|0\rangle\langle0| + \sin^2{\theta}|1\rangle\langle1|,\quad x=a,\,b.
\end{align*}
By direct calcution, we have
\begin{align*}
&I(|\psi^{\rm out}_\theta\rangle)=-4\left(\cos^2{\theta}\log\cos{\theta}+\sin^2{\theta}\log\sin{\theta}\right),\\
&\mathcal{M}_{\mathrm{SRE}_2}(|\psi^{\rm out}_\theta\rangle)=\log\!\left(
\frac{1 + f(\theta)^{2}+ g(\theta)^{2}+ \frac{3}{2}\sin^2(2\theta)}{
1 + f(\theta)^{4}+ g(\theta)^{4} + \frac{3}{8}\sin^4(2\theta)}
\right),\\
&\mathcal{M}_{L^1}\!\left(|\psi^{\mathrm{out}}_\theta\rangle\right)
= -2\log\!\left(1 + f(\theta)+ g(\theta)\right)\nonumber\\
&\qquad\qquad\,\,\qquad + \log\!\left(3\!\left(1 + f(\theta)+ g(\theta)+3\big|\sin2\theta\big|\right)\right),\\
&\mathcal{M}_{\rm mana}(|\psi^{\rm out}_\theta\rangle )
=\log\Big(1+\frac{2}{3}|\sin2\theta|\Big),
\end{align*}
where $f(\theta)=|\cos^2\theta - e^{\pi {\rm i}/3}\sin^2\theta|$ and $g(\theta)=|\cos^2\theta + e^{2\pi {\rm i}/3}\sin^2\theta|$.

The variation of the four measures with the parameter~$\theta$  is shown in Fig.~\ref{fig:3}(b).
 When $\theta=0$ or $\pi/2$ ,  the input states $|\psi_{0}\rangle=|0\rangle$ and   $|\psi_{\pi/2}\rangle=|1\rangle$ are  incoherent stabilizer  states, and the  output states $|\psi^{\rm out}_{0}\rangle=|00\rangle$  and  $|\psi^{\rm out}_{\pi/2}\rangle=|11\rangle$ are  product stabilizer states; consequently, all four  quantifiers of correlations  vanish.   When $\theta=\pi/4$,  the input state $|\psi_{\pi/4}\rangle=(|0\rangle+|1\rangle)/\sqrt{2}$ possesses the greatest  $l_2$-norm coherence and  mana among all $|\psi_{\theta}\rangle$, and the corresponding output state $|\psi^{\rm out}_{\pi/4}\rangle$  of  beamsplitter CSUM$_3$  achieves the maximum quantum mutual information, mutual  $L^1$-norm magic,  mutual stabilizer 2-Rényi entropy,  and mutual mana.

In Table ~\ref{tab1}, we compare the amount of  correlations: quantum mutual information $I(\rho_{\phi,p}^{\rm out})$,  mutual $L^1$-norm magic  $\mathcal{M}_{L^1}(\rho_{\phi,p}^{\rm out})$,  mutual stabilizer 2-Rényi entropy $\mathcal{M}_{\mathrm{SRE}_2}(\rho_{\phi,p}^{\rm out})$,  and mutual mana $\mathcal{M}_{\rm mana}(\rho_{\phi,p}^{\mathrm{out}})$ generated by passing the noisy input state  $\rho_{\phi,p}=p|\phi\rangle\langle\phi|+ (1-p) {\bf 1}/3$ through the  qutrit  beamsplitter $\mathrm{CSUM}_3$.  We consider four special qutrit input magic  states $|\phi\rangle$: the strange state $|S\rangle=(|1\rangle-|2\rangle)/\sqrt{2}$, the Norrell state $|N\rangle=(-|0\rangle+2|1\rangle-|2\rangle)/\sqrt{6}$, the  $T$-state $|T\rangle=(e^{{\rm i}2\pi/9}|0\rangle+|1\rangle+e^{-{\rm i}2\pi/9}|2\rangle)/\sqrt{3}$,
 and the  $H$-state $|H\rangle=\big((1+\sqrt{3})|0\rangle+|1\rangle+e^{-{\rm i}2\pi/9}|2\rangle\big)\big{/}\sqrt{2(3+\sqrt{3})}$. 

\begin{figure}[!t]
    \centering
    \begin{subfigure}{0.235\textwidth}
        \centering
        \includegraphics[width=\textwidth]{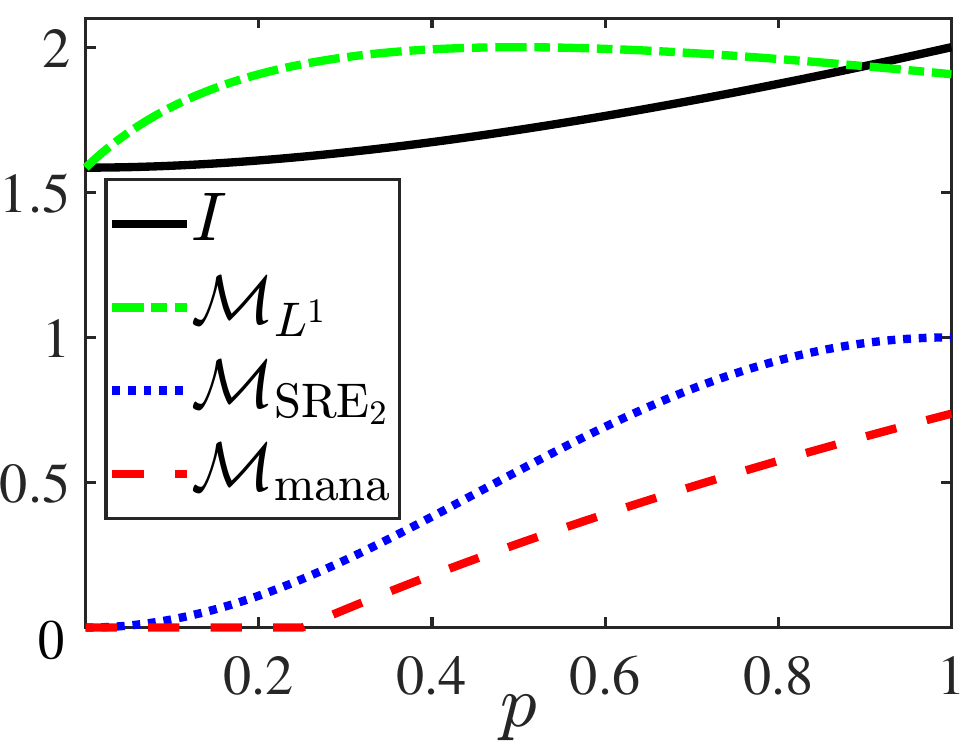}
        \caption{ $\rho_{S,p}^{\mathrm{out}}$}
        \label{fig:4a}
    \end{subfigure}
    \hfill
    \begin{subfigure}{0.235\textwidth}
        \centering
        \includegraphics[width=\textwidth]{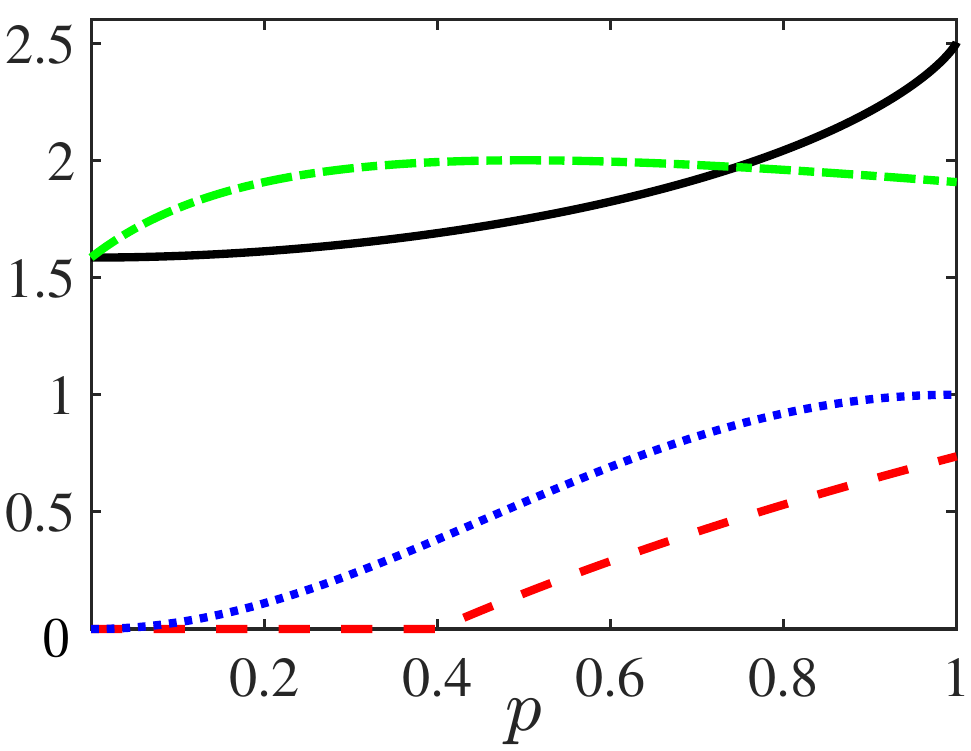}
        \caption{$\rho_{N,p}^{\mathrm{out}}$}
        \label{fig:4b}
    \end{subfigure}
    \hfill
    \begin{subfigure}{0.23512\textwidth}
        \centering
        \includegraphics[width=\textwidth]{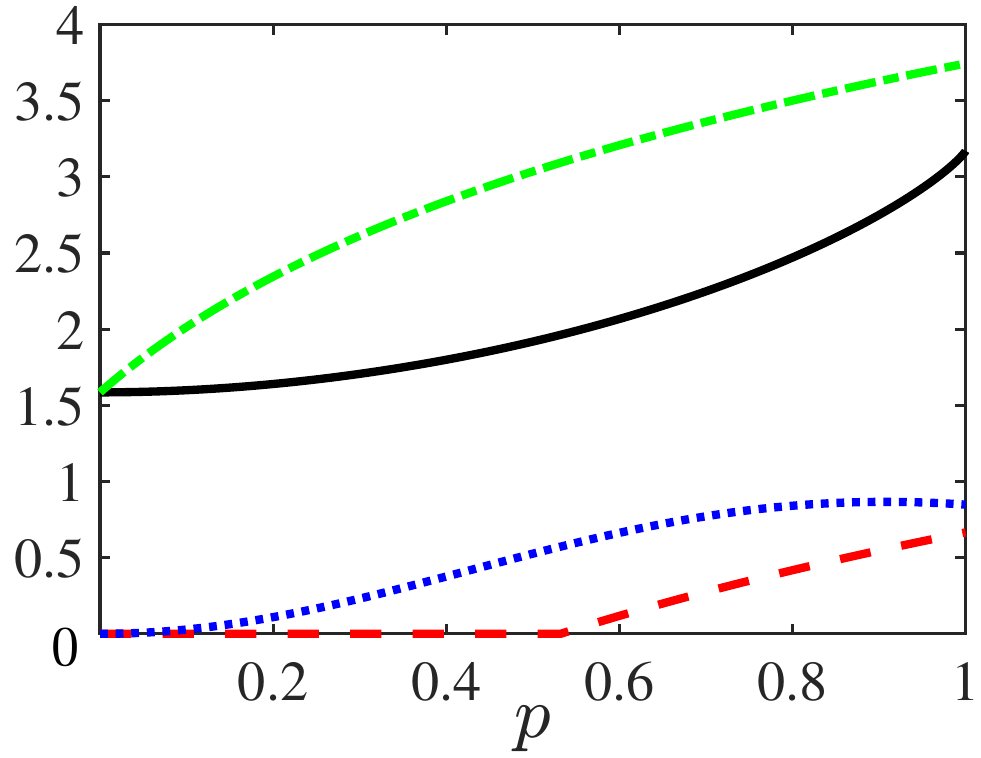}
        \caption{$\rho_{T,p}^{\mathrm{out}}$}
        \label{fig:4c}
    \end{subfigure}
    \hfill
    \begin{subfigure}{0.235\textwidth}
        \centering
        \includegraphics[width=\textwidth]{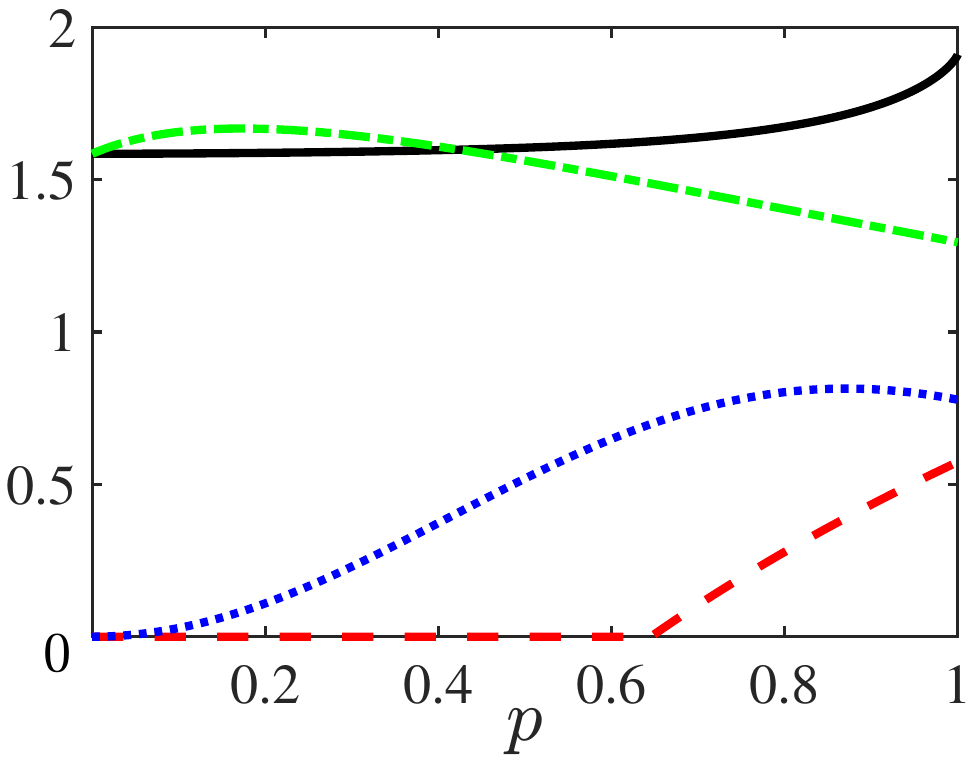}
        \caption{$\rho_{H,p}^{\mathrm{out}}$}
        \label{fig:4d}
    \end{subfigure}
    \caption{Comparison of  quantum mutual information $I(\rho_{\phi,p}^{\rm out})$ (black solid line), mutual $L^1$-norm magic  $\mathcal{M}_{L^1}(\rho_{\phi,p}^{\rm out})$ (green  dash-dotted line), mutual stabilizer 2-Rényi entropy $\mathcal{M}_{\mathrm{SRE}_2}(\rho_{\phi,p}^{\rm out})$ (blue dotted line),  and mutual mana $\mathcal{M}_{\rm mana}(\rho_{\phi,p}^{\mathrm{out}})$ (red dashed line) versus the noise  parameter~$p\in[0,1]$  for the  output states: (a) $\rho_{S,p}^{\mathrm{out}}$, (b) $\rho_{N,p}^{\mathrm{out}}$, (c) $\rho_{T,p}^{\mathrm{out}}$, (d) $\rho_{H,p}^{\mathrm{out}}$.
   All output states $\rho_{\phi,p}^{\rm out}=\mathrm{CSUM}_3(\rho_{\phi,p}\otimes|0\rangle\langle 0|)\mathrm{CSUM}^{\dag}_3$ are generated by passing the noisy input state      $\rho_{\phi,p}=p|\phi\rangle\langle\phi|+ (1-p) {\bf 1}/3$ through the  qutrit  beamsplitter $\mathrm{CSUM}_3$.
    }
    \label{fig:4}
\end{figure}

To illustrate how these quantities vary with the  noise parameter $p$, the behaviors of the four correlations are depicted in Fig.~\ref{fig:4} for the four input qutrit magic states $\{|S\rangle, |N\rangle, |T\rangle, |H\rangle\}$. As observed from the figure, the quantum mutual information $I(\rho_{\phi,p}^{\rm out})$ increases monotonically in the four cases, starting from $\log 3$ at $p=0$ and growing steadily with the input state purity. 
In contrast, the mutual $L^1$-norm magic $\mathcal{M}_{L^1}(\rho_{\phi,p}^{\rm out})$ exhibits clear state-dependent behavior: while it also begins at $\log 3$ for $p = 0$, it follows a non-monotonic trend which first increases and then decreases for $|\phi\rangle=|S\rangle$, $|N\rangle$, and $|H\rangle$, yet increases monotonically for $|\phi\rangle=|T\rangle$. This marked difference underscores the sensitivity of $\mathcal{M}_{L^1}(\rho_{\phi,p}^{\rm out})$ not only to noise but also to the intrinsic structure of the input state. Both the mutual stabilizer 2-Rényi entropy $\mathcal{M}_{\mathrm{SRE}_2}(\rho_{\phi,p}^{\rm out})$ and the mutual mana $\mathcal{M}_{\mathrm{mana}}(\rho_{\phi,p}^{\rm out})$ increase monotonically for the four states and vanish at $p = 0$. Notably, $\mathcal{M}_{\mathrm{mana}}(\rho_{\phi,p}^{\rm out})$ remains zero over the interval $0 < p < p_{\mathrm{crit}}(\phi)$, specifically
\begin{align*}
 & p_{\mathrm{crit}}(S)=\frac{1}{4},\quad p_{\mathrm{crit}}(T)=\frac{1}{2\cos{(\pi/9)}}\approx0.53,\\  
  &p_{\mathrm{crit}}(N)=\frac{2}{5},\quad p_{\mathrm{crit}}(H)=\frac{4}{1+3\sqrt{3}}\approx0.65.
\end{align*}
 $\mathcal{M}_{\mathrm{mana}}(\rho_{\phi,p}^{\rm out})$  begins to rise only after $p$ surpasses the critical value $p_{\mathrm{crit}}$, which constitutes a clear signature of threshold behavior in the establishment of magic correlations. 
 Although the four  quantifiers  of correlations  share many common traits, they also exhibit striking differences. In general, they yield different orderings of  (magic) correlations.

\section{Summary and Discussion}

 In this work, we have introduced mutual mana as a measure of magic correlations between subsystems, motivated by its additivity and structural analogy to quantum mutual information.  While related quantities have appeared under various names, such as global magic in Ref.~\cite{Sarkar2020}, the connected component of mana in Ref.~\cite{White2021}, and long-range mana in Ref.~\cite{Fliss2021}, we have investigated its role in characterizing and converting   local magic into correlations under discrete  beamsplitters.

We have shown that mutual mana is a well-defined correlation measure that vanishes for product states and  remains invariant under local Clifford operations. 
 Furthermore,  we have demonstrated that discrete beamsplitters provide a natural platform for generating magic correlations. Specifically, the coupling between a magic state and a stabilizer state results in the complete conversion of local magic into mutual mana. This fundamental phenomenon, rooted in the Clifford nature of beamsplitters which conserves total  magic resource, reveals a mechanism for redistributing magic resources into quantum correlations and provides an operational interpretation of mana.

We have established analytical bounds for the mutual mana of maximally coherent states, with the maximum being bounded by $(\log d)/2$, and have demonstrated that this bound is tight in dimensions 3 and 5.  We have derived explicit analytical expressions for the mutual mana of the output states generated by a discrete beamsplitter, with the input being the  product of the stabilizer state $|0\rangle$ with typical qutrit states, such as noisy real pure states and maximally coherent states.  We have illustrated the dependence of mutual mana on the state parameters and identified the conditions for its maximization. The study of mutual mana in higher-dimensional and many-body systems represents a compelling frontier for further exploration.

Through comparative analysis of several examples with other correlation measures, such as quantum mutual information, mutual stabilizer 2-Rényi entropy, and mutual $L^1$-norm magic, we have shown that these measures capture distinct aspects of (magic) correlations and exhibit particular merits in different physical contexts. Numerical examples highlight both similarities and differences in their behavior, emphasizing the unique features of mutual mana. A more detailed comparative study of different magic correlations  is also desirable.

Several interesting directions for future research emerge. One natural direction is to investigate a measure of magic correlations that is invariant under local unitary transformations. In this context, we define the nonlocal mana of a bipartite quantum state $\rho_{ab}$ as
\begin{align*}
\mathcal{M}_{\mathrm{non}}(\rho_{ab}):= \min_{{ U_a\otimes U_b}} \mathrm{Mana}((U_a\otimes U_b)\rho_{ab}(U_a\otimes U_b)^\dagger),  
\end{align*}
where $U_a$ and $U_b$ are local unitary operators on $\mathcal{H}_a$ and $\mathcal{H}_b$, respectively. The nonlocal mana vanishes for both product and stabilizer states, is invariant under local unitary transformations, and is subadditive under tensor products of independent states.  Formal proofs are given  in Appendix~\ref{nonlocalmana}. Exploring the theoretical implications and experimental applications of nonlocal mana and mutual mana is a promising direction for future work.

Mutual mana provides a valuable addition to the characterization of quantum resources by offering a tailored measure for magic correlations. Its mathematical properties and realization through discrete beamsplitters make it both theoretically appealing and practically relevant for understanding and  harnessing magic  resources in quantum technologies.

 The necessary conditions for the vanishing of mutual mana warrant further investigation. Exploring magic correlations via broader magic-theoretic frameworks and applying these measures to quantum interference, quantum error correction, and  quantum simulation represent promising avenues for future work. We expect that this research will provide crucial insights for the architecture of universal quantum computation and information processing.

\section*{ACKNOWLEDGMENTS}
This work was supported by  the Shuimu  Tsinghua
Scholar Program of Tsinghua University, Grant No. 2024SM378 and  the Postdoctoral Science Foundation of China, Grant No.  GZB20250711.

\vskip 0.3cm

\appendix

\section{Proof of Proposition 1}\label{proof_prop1}

Recall that any quantum state $\rho$ can be reconstructed from its discrete Wigner function \cite{Veitch2014,Wang2019}
\begin{equation}
    \rho = \sum_{k,l \in \mathbb{Z}_d} W_\rho(k,l)\, A_{k,l}, \nonumber
\end{equation}
where $\{A_{k,l}\}_{k,l \in \mathbb{Z}_d}$ are the phase-space point operators. These operators form a Hermitian, trace-one, orthogonal, and complete operator basis as  shown in Eqs. (\ref{Aklpro1}) and (\ref{Aklpro2}), 

Consequently, the purity of $\rho$ admits the following representation in terms of its Wigner function:
\begin{align*}
\mathrm{tr}(\rho^2) = \sum_{k,l \in \mathbb{Z}_d} d\, |W_\rho(k,l)|^2.
\end{align*}
Applying the Cauchy--Schwarz inequality, we obtain
\begin{align*}
\sum_{k,l \in \mathbb{Z}_d} |W_\rho(k,l)| 
\leq& \sqrt{d^2}\,\sqrt{\sum_{k,l \in \mathbb{Z}_d} |W_\rho(k,l)|^2} \nonumber\\
=& d \cdot \sqrt{\tfrac{1}{d}\,\mathrm{tr}(\rho^2)}  \nonumber\\
=& \sqrt{d\, \mathrm{tr}(\rho^2)},
\end{align*}
which implies that
\begin{align}
\mathrm{Mana}(\rho)
= \log\!\Big(\sum_{k,l \in \mathbb{Z}_d} |W_\rho(k,l)|\Big) 
\leq \tfrac{1}{2}\log\!\bigl(d\, \mathrm{tr}(\rho^2)\bigr). \nonumber  
\end{align}

\section{Proof of Proposition 2}\label{proof_prop2}
 
For item (1), recall from Proposition 2 of Ref.~\cite{Feng2025}  that for any discrete Heisenberg--Weyl operators $D_{k_1,l_1}$ and $ D_{k_2,l_2}$,
\begin{align}
    B_G\,(D_{k_1,l_1}\otimes D_{k_2,l_2})\,B_G^\dagger
    = D_{a_1,b_1}\otimes D_{a_2,b_2},  \nonumber
\end{align}
where
\begin{align*}
    a_1=&g(\delta k_1-\gamma k_2),\quad  b_1=\alpha l_1+\beta l_2,  \\
    a_2=&g(\alpha k_2-\beta k_1), \quad  b_2=\delta l_2+\gamma l_1,
\end{align*}
with all indices  taken modulo~\(d\).
Using the definition \(A_{k,l}=D_{k,l}A_{0,0}D_{k,l}^\dagger\), we have
\begin{align*}
    &B_G\bigl(A_{k_1,l_1}\otimes A_{k_2,l_2}\bigr)B_G^\dagger \nonumber\\
    =&B_G\bigl((D_{k_1,l_1}\otimes D_{k_2,l_2})(A_{0,0}\otimes A_{0,0})
    (D_{k_1,l_1}^\dagger\otimes D_{k_2,l_2}^\dagger)\bigr)B_G^\dagger \nonumber\\
    =&(D_{a_1,b_1}\otimes D_{a_2,b_2})\,
      \bigl(B_G(A_{0,0}\otimes A_{0,0})B_G^\dagger\bigr)\,
      (D_{a_1,b_1}^\dagger\otimes D_{a_2,b_2}^\dagger).
\end{align*}
It remains to verify that \(A_{0,0}\otimes A_{0,0}\) is invariant under \(B_G\).
Since
\begin{align*}
  A_{0,0}\otimes A_{0,0}
  = \frac{1}{d^2}\!\sum_{k_1,l_1,k_2,l_2\in\mathbb{Z}_d}
  D_{k_1,l_1}\otimes D_{k_2,l_2},
\end{align*}
we obtain
\begin{align*}
  B_G(A_{0,0}\otimes A_{0,0})B_G^\dagger 
  =& \frac{1}{d^2}\!\sum_{k_1,l_1,k_2,l_2}
     B_G\bigl(D_{k_1,l_1}\otimes D_{k_2,l_2}\bigr)B_G^\dagger \nonumber \\
  =& \frac{1}{d^2}\!\sum_{k_1,l_1,k_2,l_2}
     D_{a_1,b_1}\otimes D_{a_2,b_2},
\end{align*}
where $(a_1,b_1,a_2,b_2)=(k_1,l_1,k_2,l_2)\, M$ with
\begin{align*}
M :=
\begin{pmatrix}
g\delta & 0 &-g\beta & 0 \\
0 & \alpha & 0 & \gamma\\
-g\gamma & 0 & g\alpha & 0 \\
0 & \beta & 0 & \delta
\end{pmatrix}.
\end{align*}
Given that \(g:=(\det G)^{-1}\pmod d\), we have 
\begin{align}
  \det M= g^2\,(\alpha\delta-\beta\gamma)^2
     = g^2\,(\det G)^2
     = 1 ,  \nonumber
\end{align}
which implies that the map \((k_1,l_1,k_2,l_2)\mapsto(a_1,b_1,a_2,b_2)\) is a bijection of \(\mathbb{Z}_d^4\).  Therefore,
\begin{align*}
  B_G(A_{0,0}\otimes A_{0,0})B_G^\dagger 
  = &\frac{1}{d^2}\sum_{k_1,l_1,k_2,l_2} D_{a_1,b_1}\otimes D_{a_2,b_2} \nonumber\\
  = &\frac{1}{d^2}\sum_{a_1,b_1,a_2,b_2} D_{a_1,b_1}\otimes D_{a_2,b_2}\nonumber\\
  =& A_{0,0}\otimes A_{0,0}.
\end{align*}
Consequently,
\begin{align*}
  &B_G\bigl(A_{k_1,l_1}\otimes A_{k_2,l_2}\bigr)B_G^\dagger \nonumber\\
  =& \bigl(D_{a_1,b_1}A_{0,0}D_{a_1,b_1}^\dagger\bigr)
     \otimes
     \bigl(D_{a_2,b_2}A_{0,0}D_{a_2,b_2}^\dagger\bigr) \nonumber\\
  =& A_{a_1,b_1}\otimes A_{a_2,b_2} \nonumber\\
  =& A_{\,g(\delta k_1-\gamma k_2),\,\alpha l_1+\beta l_2}
     \otimes
     A_{\,g(\alpha k_2-\beta k_1),\,\delta l_2+\gamma l_1}.
\end{align*}

For item (ii), we first prove that
\begin{align*}
&B_G^\dagger(A_{k,l}\otimes {\bf1})B_G  \nonumber\\
=& \frac{1}{d} \sum_{m,n\in \mathbb{Z}_d} \tau^{(l - k)(m+n)} (D_{\alpha m, g \delta n} \otimes D_{\beta m, -g \gamma n}).
\end{align*}
Using the fact $D_{k,l} D_{s,t} = \tau^{\,ls - kt} D_{k+s,\,l+t}$,  it can be verified that
\begin{align*}
D_{k,l} D_{m,n} D_{k,l}^\dagger 
=& \tau^{\,lm - kn}\, D_{k+m,\,l+n} D_{-k,-l}\nonumber\\
=& \tau^{\,lm - kn}\, \tau^{-(l+n)k + (k+m)l} D_{m,n}\nonumber\\
=& \tau^{\,2(lm - kn)} D_{m,n} \nonumber\\
=&\omega^{\,lm - kn}D_{m,n}.
\end{align*}
Therefore,
\begin{align}
A_{k,l}
=&D_{k,l}A_{0,0}D_{k,l}^\dagger \nonumber\\
=&\frac{1}{d}\sum_{m,n\in \mathbb{Z}_d} D_{k,l}D_{m,n}D_{k,l}^\dagger \nonumber\\
=&\frac{1}{d}\sum_{m,n\in \mathbb{Z}_d}\omega^{\,lm - kn} D_{m,n}. \label{Akl}
\end{align} 
Furthermore, combining Eq. \eqref{Akl} and  the following covariance relation established in Eq.~(C5) of Ref.~\cite{Feng2025},
\begin{equation*}
B_G^\dagger(D_{k,l}\otimes \mathbf{1})B_G \;=\; D_{\alpha k,\,g\delta l}\otimes D_{\beta k,\,-g\gamma l},
\end{equation*}
we obtain 
\begin{align*}
&B_G^\dagger (A_{k,l} \otimes \mathbf{1}) B_G\nonumber\\
=& \frac{1}{d} \sum_{m,n\in \mathbb{Z}_d}\omega^{\,lm - kn} B_G^\dagger (D_{m,n} \otimes \mathbf{1}) B_G  \nonumber\\
=& \frac{1}{d} \sum_{m,n\in \mathbb{Z}_d} \omega^{\,lm - kn} (D_{\alpha m, g \delta n} \otimes D_{\beta m, -g \gamma n}).
\end{align*}

We next prove that
\begin{align*}
&B_G^\dagger ({\bf1} \otimes A_{k,l}) B_G \nonumber\\
=& \frac{1}{d} \sum_{m,n \in \mathbb{Z}_d} \omega^{l m - k n} 
(D_{-\gamma m, g \beta n} \otimes D_{-\delta m, -g \alpha n}).
\end{align*}
Using  the swap operator $S$,  we obtain
\begin{align*}
B_G^\dagger (\mathbf{1} \otimes A_{k,l}) B_G = B_G^\dagger S (A_{k,l} \otimes \mathbf{1}) S B_G.
\end{align*}
Moreover, it is easy to verify that there exists a matrix $G' = \begin{pmatrix} -\gamma & -\delta \\ -\alpha & -\beta \end{pmatrix}$ such that $S B_G = B_{G'}$, where
\begin{align*}
&\alpha' = -\gamma, \quad \beta' = -\delta, \quad\gamma' = -\alpha,  \quad \delta' = -\beta, \\
&g' = \det G' = \gamma\beta - \alpha\delta = -g.
\end{align*}
Therefore,
\begin{align*}
B_G^\dagger (\mathbf{1} \otimes A_{k,l}) B_G 
= B_G^\dagger S (A_{k,l} \otimes \mathbf{1}) S B_G
= B_{G'}^\dagger (A_{k,l} \otimes \mathbf{1}) B_{G'}.
\end{align*}
From  Eq.~(\ref{BGAkl1}), for any matrix 
$G = \begin{pmatrix} \alpha & \beta \\ \gamma & \delta \end{pmatrix}$,  we have
\begin{align*}
B_G^\dagger (A_{k,l} \otimes \mathbf{1}) B_G 
= \frac{1}{d} \sum_{m,n \in \mathbb{Z}_d} \omega^{l m - k n} 
(D_{\alpha m, g \delta n} \otimes D_{\beta m, -g \gamma n}).
\end{align*}
Therefore,
\begin{align*}
&B_G^\dagger (\mathbf{1} \otimes A_{k,l}) B_G \nonumber\\
=&B_{G'}^\dagger (A_{k,l} \otimes \mathbf{1}) B_{G'} \nonumber\\
=& \frac{1}{d} \sum_{m,n \in \mathbb{Z}_d} \omega^{l m - k n} 
(D_{\alpha' m, g' \delta' n} \otimes D_{\beta' m, -g' \gamma' n}) \nonumber \\
=& \frac{1}{d} \sum_{m,n \in \mathbb{Z}_d} \omega^{l m - k n} 
(D_{-\gamma m, (-g)(-\beta) n} \otimes D_{-\delta m, -(-g)(-\alpha) n}) \nonumber \\
=& \frac{1}{d} \sum_{m,n \in \mathbb{Z}_d} \omega^{l m - k n} 
(D_{-\gamma m, g \beta n} \otimes D_{-\delta m, -g \alpha n}).
\end{align*}

\section{Proof of Proposition 3}\label{proof_prop3}

We first prove that
\begin{align}
    \operatorname{tr}\Big((\rho \otimes |0\rangle\langle 0|) B_G^\dagger (A_{k,l} \otimes \mathbf{1}) B_G \Big) = \rho_{j_0 j_0}, \nonumber
\end{align}
where $j_0 =k(g\delta)^{-1}= k{\delta}^{-1}\det G  \quad  (\mathrm{mod}\ d)$. 

From Eq. (\ref{BGAkl1}),  we have
\begin{align*}
&\operatorname{tr}\Big((\rho \otimes |0\rangle\langle 0|) B_G^\dagger (A_{k,l} \otimes \mathbf{1}) B_G \Big) \nonumber\\
=& \frac{1}{d} \sum_{m,n} \omega^{l m - k n} \operatorname{tr}\Big((\rho \otimes |0\rangle\langle 0|) (D_{\alpha m, g \delta n} \otimes D_{\beta m, - g \gamma n})\Big),\nonumber\\
=&\frac{1}{d} \sum_{m,n} \omega^{l m - k n} \operatorname{tr}(\rho D_{\alpha m, g \delta n}) \operatorname{tr}(|0\rangle\langle 0| D_{\beta m, - g \gamma n}).
\end{align*}
Recall that 
\begin{align*}
\operatorname{tr}(|0\rangle\langle 0| D_{\beta m, - g \gamma n}) 
=& \tau^{-\beta g \gamma m n} \langle 0 | X^{\beta m} Z^{- g \gamma n}|0\rangle \nonumber\\
=&\tau^{-\beta g \gamma m n} \langle 0 |X^{\beta m} |0\rangle \nonumber\\
=&\tau^{-\beta g \gamma m n}\langle 0 |\beta m\rangle\nonumber\\
=& \tau^{- \beta g \gamma m n} \delta_{\beta m, 0}.
\end{align*}
Since $\beta \neq 0$, this requires $m = 0$, and thus
\begin{align}
\operatorname{tr}(|0\rangle\langle 0| D_{\beta m, - g \gamma n}) =\langle 0 | D_{\beta m, - g \gamma n} | 0 \rangle= \delta_{m,0}. \nonumber
\end{align}
Therefore,
\begin{align}
\operatorname{tr}\Big((\rho \otimes |0\rangle\langle 0|) B_G^\dagger (A_{k,l} \otimes \mathbf{1}) B_G \Big) 
= \frac{1}{d} \sum_{n} \omega^{- k n} \operatorname{tr}(\rho Z^{g \delta n}). \nonumber
\end{align}
Writing $\operatorname{tr}(\rho Z^{g \delta n})$ in the computational basis:
\begin{align}
\operatorname{tr}(\rho Z^{g \delta n}) = \sum_{j \in \mathbb{Z}_d} \rho_{jj} \, \omega^{g \delta n j},   \nonumber
\end{align}
where $\rho_{jj}=\langle j|\rho|j\rangle,$
we obtain
\begin{align*}
&\operatorname{tr}\Big((\rho \otimes |0\rangle\langle 0|) B_G^\dagger (A_{k,l} \otimes \mathbf{1}) B_G \Big) \nonumber\\
=&\frac{1}{d} \sum_{n} \omega^{- k n} \sum_{j} \rho_{jj} \, \omega^{g \delta n j}\nonumber\\
=& \frac{1}{d} \sum_{j} \rho_{jj} \sum_{n} \omega^{n (g \delta j - k)}.
\end{align*}
Note that the  sum $\sum_{n=0}^{d-1} \omega^{n a}$ equals $d$ if $a \equiv 0 \mod d$, and $0$ otherwise. Therefore, 
\begin{align*}
\operatorname{tr}\big((\rho \otimes |0\rangle\langle 0|) B_G^\dagger (A_{k,l} \otimes \mathbf{1}) B_G \big)
=\frac{1}{d} \sum_{j} \rho_{jj} \cdot d \cdot \delta_{j, j_0}
= \rho_{j_0 j_0},
\end{align*}
where $j_0 := k (g \delta)^{-1} = k{\delta}^{-1} \det G \quad ({\rm mod} \ \ d).$

We next prove that
\begin{align}
\operatorname{tr}\Big((\rho \otimes |0\rangle\langle 0|) B_G^\dagger ({\bf1} \otimes A_{k,l}) B_G \Big) = \rho_{j_1 j_1},  \nonumber
\end{align}
where $j_1= k\,(g\beta)^{-1}=k {\beta}^{-1}\det G \quad ({\rm mod} \, \ d).$\\
From Eq. (\ref{BGAkl2}), we have 
\begin{align*}
&\operatorname{tr}\!\Big((\rho\otimes |0\rangle\!\langle 0|)\, B_G^\dagger ({\bf1}\otimes A_{k,l}) B_G \Big) \nonumber\\
=& \frac{1}{d}\sum_{m,n}\omega^{lm-kn}\,
\operatorname{tr}\!\big(\rho\, D_{-\gamma m,\,g\beta n}\big)\,
\langle 0|D_{-\delta m,\,-g\alpha n}|0\rangle.
\end{align*}
Note that
\begin{align*}
\langle 0|D_{-\delta m,\,-g\alpha n}|0\rangle
= \tau^{\delta m g\alpha n}\,\langle 0|-\delta m\rangle
= \tau^{\delta m g\alpha n}\,\delta_{\delta m,0}.
\end{align*}
Since $\delta \neq 0$, this requires $m = 0$, and thus
\begin{align*}
\operatorname{tr}(|0\rangle\langle 0| D_{-\delta m,\,-g\alpha n}) =\langle 0 | D_{-\delta m,\,-g\alpha n} | 0 \rangle= \delta_{m,0}.
\end{align*}
Therefore,
\begin{align*}
&\operatorname{tr}\!\Big((\rho\otimes |0\rangle\!\langle 0|)\, B_G^\dagger ({\bf1}\otimes A_{k,l}) B_G \Big)  \nonumber\\
=& \frac{1}{d}\sum_{n}\omega^{-kn}\,\operatorname{tr}\!\big(\rho\, Z^{g\beta n}\big)\nonumber\\
=& \frac{1}{d}\sum_{n}\omega^{-kn}\sum_{j}\rho_{jj}\,\omega^{g\beta n\, j}\nonumber\\
=& \frac{1}{d}\sum_{j}\rho_{jj} \sum_{n}\omega^{n(g\beta j-k)}\nonumber\\
=& \sum_{j}\rho_{jj}\,\delta_{g\beta j,\,k} \nonumber\\
=& \rho_{j_1 j_1},
\end{align*}
with $j_1\equiv k\,(g\beta)^{-1} \quad ({\rm mod} \,\ d)$.

\section{Proof of Proposition 4}  \label{proof_prop4}

Item (1) follows directly from the definition of $\mathcal{M}_{\rm mana}(\rho_{a}\otimes\rho_b)$ and  Eq. (\ref{additivity}).

For item (2), for any Clifford operators $C_1$ and $C_2$, and define
\begin{align}
\sigma_{ab} = (C_1 \otimes C_2)\rho_{ab}(C_1 \otimes C_2)^\dagger, \nonumber
\end{align}
with reduced states
\begin{align}
\sigma_a = \mathrm{tr}_b(\sigma_{ab}), \qquad
\sigma_b = \mathrm{tr}_a(\sigma_{ab}). \nonumber
\end{align}
The corresponding mutual mana is  given by
\begin{align}
    \mathcal{M}_{\rm mana}(\sigma_{ab})
    = \mathrm{Mana}(\sigma_{ab}) - \mathrm{Mana}(\sigma_a) - \mathrm{Mana}(\sigma_b). \nonumber
\end{align}
Tracing out the subsystems gives
\begin{align*}
    \sigma_a 
        &= C_1\,\mathrm{tr}_b(\rho_{ab})\,C_1^\dagger 
         = C_1 \rho_a C_1^\dagger, \\
    \sigma_b 
        &= C_2\,\mathrm{tr}_a(\rho_{ab})\,C_2^\dagger 
         = C_2 \rho_b C_2^\dagger.
\end{align*}
Using Eq.~\eqref{invariance}, it follows that
\begin{align*}
    \mathcal{M}_{\rm mana}(\sigma_{ab})
    &= \mathrm{Mana}(\sigma_{ab}) - \mathrm{Mana}(\sigma_a) - \mathrm{Mana}(\sigma_b) \\
    &= \mathrm{Mana}(\rho_{ab}) - \mathrm{Mana}(\rho_a) - \mathrm{Mana}(\rho_b) \\
    &= \mathcal{M}_{\rm mana}(\rho_{ab}).
\end{align*}

\section{Proof of Theorem 1}\label{proof_thm}

Let $\rho_{ab}^{\rm out} := B_G \bigl(\rho \otimes |0\rangle\langle 0|\bigr) B_G^\dagger$ and denote its reduced states by
\begin{align}
    \rho_a^{\rm out} = \mathrm{tr}_b(\rho_{ab}^{\rm out}), \qquad
    \rho_b^{\rm out} = \mathrm{tr}_a(\rho_{ab}^{\rm out}).  \nonumber
\end{align}
The corresponding mutual mana is then given by
\begin{align}
    \mathcal{M}_{\rm mana}(\rho_{ab}^{\rm out})
    = \mathrm{Mana}(\rho_{ab}^{\rm out}) - \mathrm{Mana}(\rho_a^{\rm out}) - \mathrm{Mana}(\rho_b^{\rm out
    }). \nonumber
\end{align}
Since $B_G$ is a Clifford operator, and using Eqs.~\eqref{invariance} and \eqref{additivity}, we have
\begin{align}
    \mathrm{Mana}(\rho_{ab}^{\rm out})
    = \mathrm{Mana}\bigl(\rho \otimes |0\rangle\langle 0|\bigr) 
    = \mathrm{Mana}(\rho). \nonumber
\end{align}
For the reduced states $\rho_a^{\rm out}$ and $\rho_b^{\rm out}$, we have
\begin{align*}
&\mathrm{Mana}(\rho_a^{\rm out}) \nonumber\\
=& \log \sum_{k,l \in \mathbb{Z}_d} \bigl| W_{\rho_a^{\rm out}}(k,l) \bigr|\nonumber\\
=&\log \sum_{k,l \in \mathbb{Z}_d} \bigl|\frac{1}{d} {\rm tr}(\rho_a^{\rm out}A_{k,l}) \bigr| \nonumber\\
=&\log \sum_{k,l \in \mathbb{Z}_d} \bigl| \frac{1}{d} {\rm tr}(\rho_{ab}^{\rm out}(A_{k,l}\otimes{\bf1})) \bigr|  \nonumber  \\
=&\log \sum_{k,l \in \mathbb{Z}_d} \bigl| \frac{1}{d} {\rm tr}((\rho \otimes |0\rangle\langle 0|) B_G^\dagger(A_{k,l}\otimes{\bf1}) B_G) \bigr|,
\end{align*}
and 
\begin{align*}
&\mathrm{Mana}(\rho_b^{\rm out}) \nonumber\\
=& \log \sum_{k,l \in \mathbb{Z}_d} \bigl| W_{\rho_b^{\rm out}}(k,l) \bigr|\nonumber\\
=&\log \sum_{k,l \in \mathbb{Z}_d} \bigl|\frac{1}{d} {\rm tr}(\rho_b^{\rm out}A_{k,l}) \bigr| \nonumber\\
=&\log \sum_{k,l \in \mathbb{Z}_d} \bigl| \frac{1}{d} {\rm tr}(\rho_{ab}^{\rm out}({\bf1}\otimes A_{k,l})) \bigr|  \nonumber  \\
=&\log \sum_{k,l \in \mathbb{Z}_d} \bigl| \frac{1}{d} {\rm tr}((\rho \otimes |0\rangle\langle 0|) B_G^\dagger({\bf1}\otimes A_{k,l}) B_G) \bigr|.
\end{align*}
Then, combining  Proposition \ref{prop3}, we obtain
\begin{align*}
\mathrm{Mana}(\rho_a^{\rm out})
=&\log \sum_{k,l \in \mathbb{Z}_d} \bigl| \frac{1}{d} \rho_{j_0 j_0} \bigr|
=\log \sum_{k \in \mathbb{Z}_d}  \rho_{j_0 j_0}
=\log1
=0,\\
\mathrm{Mana}(\rho_b^{\rm out})
=&\log \sum_{k,l \in \mathbb{Z}_d} \bigl| \frac{1}{d} \rho_{j_1 j_1} \bigr|
=\log \sum_{k \in \mathbb{Z}_d}  \rho_{j_1 j_1}
=\log1
=0.
 \end{align*}
Therefore, the mutual mana of $\rho_{ab}^{\rm out}$ reduces to
\begin{align*}
    \mathcal{M}_{\rm mana}(\rho_{ab}^{\rm out}) 
    &= \mathrm{Mana}(\rho_{ab}^{\rm out}) - \mathrm{Mana}(\rho_a^{\rm out}) - \mathrm{Mana}(\rho_b^{\rm out}) \nonumber\\
    &= \mathrm{Mana}(\rho_{ab}^{\rm out}) \nonumber\\
    &= \mathrm{Mana}(\rho).
\end{align*}

\section{Proof of Proposition 5}\label{proof_prop5}

As shown in Ref.~\cite{Feng2025}, the reduced states $\rho_a^{\rm out} = \operatorname{tr}_b \rho_{ab}^{\rm out}$ and $\rho_b^{\rm out} = \operatorname{tr}_a \rho_{ab}^{\rm out}$ of the output $\rho_{ab}^{\rm out}=B_G(|\psi_{\boldsymbol{\theta}}\rangle\otimes|0\rangle)(\langle\psi_{\boldsymbol{\theta}}|\otimes\langle0|)B_G^\dagger$ are both maximally mixed,
\begin{align}
\rho_a^{\rm out} = \rho_b^{\rm out} = \frac{\mathbf{1}}{d}. \nonumber
\end{align}
Combining Eqs. \eqref{invariance} and \eqref{additivity}, and the fact that  the discrete  beamsplitter $B_G$ is a Clifford operator, we obtain
\begin{align}
\mathrm{Mana}(\rho_{ab}^{\rm out}) = \mathrm{Mana}(|\psi_{\boldsymbol{\theta}}\rangle\otimes |0\rangle) =  \mathrm{Mana}(|\psi_{\boldsymbol{\theta}}\rangle).  \nonumber
\end{align}
Consequently, 
\begin{align*}
\mathcal{M}_{\rm mana}(\rho_{ab}^{\rm out}) 
= & \mathrm{Mana}(\rho_{ab}^{\rm out}) -  \mathrm{Mana}(\rho_a^{\rm out}) -  \mathrm{Mana}(\rho_b^{\rm out})  \nonumber\\
=& \mathrm{Mana}(|\psi_{\boldsymbol{\theta}}\rangle)  \nonumber\\
\leq& \frac{1}{2}\log d,
\end{align*}
where the last inequality is derived from Proposition 1.

Therefore,
\begin{align*}
\max_{\boldsymbol{\theta} \in \mathbb{R}^{d-1}} \; \mathcal{M}_{\rm mana}\Big(B_G(|\psi_{\boldsymbol{\theta}}\rangle \otimes |0\rangle)\Big) 
=&\max_{\boldsymbol{\theta} \in \mathbb{R}^{d-1}} \mathrm{Mana}(|\psi_{\boldsymbol{\theta}}\rangle) 
\end{align*}
Moreover, the equality holds  if  $d=3$ and $5$.

\section{Analytical expressions for  $\mathcal{M}_{\mathrm{SRE}_2}(\rho^{\rm out}_{H,p})$ and $\mathcal{M}_{L^1}(\rho^{\rm out}_{H,p})$}
We consider the   noisy   $H$-state as input 
 $\rho_{H,p}=p|H\rangle\langle H|+ (1-p) {\bf 1}/3$,
where the qutrit $H$-state is $|H\rangle=\big((1+\sqrt{3})|0\rangle+|1\rangle+e^{-{\rm i}2\pi/9}|2\rangle\big)\big{/}\sqrt{2(3+\sqrt{3})}$.
The corresponding output state after the  qutrit  beamsplitter $\mathrm{CSUM}_3$  is
 $\rho_{H,p}^{\rm out}=\mathrm{CSUM}_3(\rho_{H,p}\otimes|0\rangle\langle 0|)\mathrm{CSUM}^{\dag}_3$.
The  mutual stabilizer 2-Rényi entropy $\mathcal{M}_{\mathrm{SRE}_2}(\rho_{H,p}^{\rm out})$  and mutual $L^1$-norm magic  $\mathcal{M}_{L^1}(\rho_{H,p}^{\rm out})$ for the  output state  can be evaluated as 
\begin{widetext}
\begin{align}\label{ML1H}
\mathcal{M}_{L^1}(\rho^{\rm out}_{H,p}) 
=& -2\log\Big(1 + \frac{(1+\sqrt{3})p}{2}\Big)
    +\log\Big (3+\frac{3(1+\sqrt{3})p}{2}+\frac{(3-\sqrt{3})p}{2}\big|(1+\sqrt{3})(1+e^{\frac{-{\rm i} 2\pi}{9}})+e^{\frac{{\rm i} 2\pi}{9}}\big|  \notag \\
    &\quad+\frac{(3-\sqrt{3})p}{4}(\big|(1+\sqrt{3})(1+e^{\frac{-{\rm i} 8\pi}{9}})+e^{\frac{{\rm i} 8\pi}{9}}\big|+\big|(1+\sqrt{3})(1+e^{\frac{{\rm i} 4\pi}{9}})+e^{\frac{-{\rm i} 4 \pi}{9}}\big|) \notag \\
   &\quad+\frac{(3-\sqrt{3})p}{4}(\big|(1+\sqrt{3})(e^{\frac{{\rm i} 2\pi}{9}}+e^{\frac{{\rm i} 2\pi}{3}})+e^{\frac{{\rm i} 10\pi}{9}}\big|+\big|(1+\sqrt{3})(e^{\frac{{\rm i} 2\pi}{9}}+e^{\frac{{\rm i} 4\pi}{3}})+e^{\frac{{\rm i} 4 \pi}{9}}\big|\big)  \Big),
\end{align} 
and
\begin{align}\label{MSREH}
&\mathcal{M}_{\mathrm{SRE}_2}(\rho^{\rm out}_{H,p}) \notag\\
=& \log\!\Bigg[
   2(3 + \sqrt{3})^4
   \Big(24- (-2 + \sqrt{3})\big|(e^{{\rm i}\pi/9} - e^{2\pi {\rm i}/3})(1+\sqrt{3}) - e^{{\rm i}2\pi/9} \big|^2\cdot p^2\notag \\
&\quad
      - (-2 + \sqrt{3})\big|(e^{{\rm i}5\pi/9} - e^{2\pi {\rm i}/3})(1+\sqrt{3}) + e^{{\rm i}7\pi/9} \big|^2\cdot p^2 \notag \\
&\quad
      + 2 \, \!\big(18 + \sqrt{3}- 2\sqrt{3}\cos\!\frac{\pi}{9}+ (1 + 3\sqrt{3})\cos\!\frac{2\pi}{9}+ (3\sqrt{3}-1)\sin\!\frac{\pi}{18}\big)\cdot p^2\Big)
   \Bigg] \notag \\
&\quad
- \log\!\Bigg[3\Big(576(7 + 4\sqrt{3}) + \big| (e^{{\rm i}\pi/9} - e^{2\pi {\rm i}/3})(1+\sqrt{3}) - e^{{\rm i}2\pi/9} \big|^4\cdot p^4 + \big| (e^{{\rm i}5\pi/9} - e^{2\pi {\rm i}/3})(1+\sqrt{3})+ e^{{\rm i}7\pi/9}\big|^4\cdot p^4 \notag \\[3pt]
&\qquad
+ 2 \,\big(1299 + 744\sqrt{3} - 2(146 + 85\sqrt{3})\cos\!\frac{\pi}{9} + (478 + 278\sqrt{3})\cos\!\frac{2\pi}{9}+ (382+ 224\sqrt{3})\sin\!\frac{\pi}{18} \big)\cdot p^4\Big)\Bigg].
\end{align}

\end{widetext}

\section{Properties of nonlocal mana}\label{nonlocalmana}

The nonlocal mana $\mathcal{M}_{\mathrm{non}}(\rho_{ab})$   of  a bipartite quantum state $\rho_{ab}$ satisfies the following properties.

(1) The nonlocal mana vanishes for both product states and stabilizer states, i.e.,
\begin{align}
\mathcal{M}_{\mathrm{non}}(\rho_a \otimes \rho_b) = 0,\qquad \mathcal{M}_{\mathrm{non}}(\rho_{ab}) = 0,\quad\forall \rho_{ab}\in\mathrm{STAB}. \nonumber
\end{align}

(2) The nonlocal mana is invariant under  local unitary transformation in the sense that
\begin{align}
\mathcal{M}_{\mathrm{non}}\!\big[(V_a \!\otimes\! V_b)\rho_{ab}(V_a \!\otimes\! V_b)^{\dagger}\big]
= \mathcal{M}_{\mathrm{non}}(\rho_{ab}),   \nonumber
\end{align}
for any local unitary operators  $V_A\in  \mathcal{U}(\mathcal{H}_a),\, V_B\in\mathcal{U}(\mathcal{H}_b)$.

(3) The nonlocal mana is subadditive in the sense that
\begin{align}
\mathcal{M}_{\mathrm{non}}(\rho_{ab} \!\otimes\! \sigma_{a'b'})\leq \mathcal{M}_{\mathrm{non}}(\rho_{ab})+\mathcal{M}_{\mathrm{non}}(\sigma_{a'b'}).  \nonumber
\end{align}

We now proceed to prove the properties of nonlocal mana.

For item (1), we first prove $\mathcal{M}_{\mathrm{non}}(\rho_a \otimes \rho_b)=0$.
By the definition of $\mathcal{M}_{\mathrm{non}}$ and the additivity of the mana measure [Eq.~\eqref{additivity}], we have
\begin{align*}
&\mathcal{M}_{\mathrm{non}}(\rho_a \otimes \rho_b) \nonumber\\
=& \min_{U_a \otimes U_b} \mathrm{Mana}\!\left(U_a \rho_a U_a^\dagger \otimes U_b \rho_b U_b^\dagger\right)  \nonumber\\
=& \min_{U_a,\,U_b}\big[\mathrm{Mana}(U_a \rho_a U_a^\dagger) + \mathrm{Mana}(U_b \rho_b U_b^\dagger)\big].
\end{align*}
Since any quantum state can be transformed into a stabilizer state via an appropriate unitary operation, one can always find $U_a$ and $U_b$ such that 
$\mathrm{Mana}(U_a \rho_a U_a^\dagger)
 = \mathrm{Mana}(U_b \rho_b U_b^\dagger)
 = 0$. 
Therefore, 
\begin{align*}
&\mathcal{M}_{\mathrm{non}}(\rho_a \otimes \rho_b) \nonumber\\
=& \min_{U_a}\mathrm{Mana}(U_a \rho_a U_a^\dagger) + \min_{U_b} \mathrm{Mana}(U_b \rho_b U_b^\dagger) \nonumber\\
=& 0.
\end{align*}
Next, we prove that $\mathcal{M}_{\mathrm{non}}(\rho_{ab}) = 0$ for any stabilizer state $\rho_{ab}$.  
By the definition of $\mathcal{M}_{\mathrm{non}}$ and the fact that $\mathrm{Mana}(\rho_{ab})=0$ for all stabilizer states, we obtain
\begin{align*}
\mathcal{M}_{\mathrm{non}}(\rho_{ab}) 
=& \min_{U_a\otimes U_b} \mathrm{Mana}\!\left[(U_a\otimes U_b)\rho_{ab}(U_a\otimes U_b)^\dagger \right] \nonumber\\
\le& \mathrm{Mana}(\rho_{ab})\nonumber\\
=& 0.
\end{align*}
Therefore, $\mathcal{M}_{\mathrm{non}}(\rho_{ab}) = 0$ for any stabilizer state $\rho_{ab}$.

For item (2), for any unitary operators $V_a$ and $V_b$,
\begin{align*}
&\mathcal{M}_{\mathrm{non}}\!\big[(V_a \!\otimes\! V_b)\rho_{ab}(V_a \!\otimes\! V_b)^{\dagger}\big] \nonumber\\
=& \min_{U_a \otimes U_b}\mathrm{Mana}((U_aV_a \!\otimes\! U_bV_b)\rho_{ab}(U_aV_a \!\otimes\! U_bV_b)^{\dagger}) \nonumber\\
= &n \mathcal{M}_{\mathrm{non}}(\rho_{ab}).
\end{align*}

For item (3), combining the definition of $\mathcal{M}_{\mathrm{non}}$ and the additivity of the mana measure [Eq.~\eqref{additivity}], we have
\begin{align*}
&\mathcal{M}_{\mathrm{non}}(\rho_{ab} \!\otimes\! \sigma_{a'b'}) \nonumber\\
=&\min_{U_{ab}\otimes U_{a'b'}}\mathrm{Mana}((U_{ab}\otimes U_{a'b'})(\rho_{ab} \!\otimes\! \sigma_{a'b'})(U_{ab}\otimes U_{a'b'})^\dagger) \nonumber\\
=&\min_{U_{ab}}\mathrm{Mana}(U_{ab}\rho_{ab}U_{ab}^\dagger) + \min_{U_{a'b'}} \mathrm{Mana}(U_{a'b'}\sigma_{a'b'}U_{a'b'}^\dagger) \nonumber\\
\leq&\min_{U_{a}\otimes U_b}\mathrm{Mana}((U_{a}\otimes U_b)\rho_{ab}(U_{a}\otimes U_b)^\dagger) \nonumber\\
&+ \min_{U_{a'}\otimes U_{b'}} \mathrm{Mana}((U_{a'}\otimes U_{b'})\sigma_{a'b'}(U_{a'}\otimes U_{b'})^\dagger) \nonumber\\
=& \mathcal{M}_{\mathrm{non}}(\rho_{ab})+\mathcal{M}_{\mathrm{non}}(\sigma_{a'b'}).
\end{align*}



\begin{thebibliography}{}


\bibitem{Gottesman1999} D. Gottesman and  I. Chuang,  Demonstrating the viability of universal quantum computation using
teleportation and single-qubit operations, Nature {\bf402}, 390 (1999).


\bibitem{BravyiGosset2016} S. Bravyi and D. Gosset, Improved classical simulation of quantum circuits dominated by Clifford gates,
 Phys. Rev. Lett. {\bf116}, 250501 (2016).

\bibitem{seddon2021} J. R. Seddon, B.  Regula, H. Pashayan, Y.  Ouyang, and E. T.  Campbell,  Quantifying quantum speedups: Improved classical simulation from tighter magic monotones, PRX Quantum {\bf2}, 010345 (2021).


\bibitem{liuzw2022}Z. W. Liu and  A. Winter, Many-body quantum magic, PRX Quantum  {\bf3}, 020333 (2022).

\bibitem{weif2025}F. Wei and  Z. W. Liu,  Long-range nonstabilizerness from quantum codes, orders, and correlations, arXiv:2503.04566 (2025).

\bibitem{WEIFLIU2024}F. Wei and  Z. W. Liu,  Noise robustness and threshold of many-body quantum magic,  arXiv: 2410.21215 (2024).

\bibitem{chenj2024} J. Chen, Y.  Yan,  and Y. Zhou,  Magic of quantum hypergraph states, Quantum {\bf8}, 1351 (2024).

\bibitem{Gottesman1997} D. Gottesman, Stabilizer codes and quantum error correction, Ph.D. thesis, California Institute of Technology, Pasadena, CA, 1997.

\bibitem{Gottesman1998}D. Gottesman,  The Heisenberg representation of quantum computers, arXiv:quant-ph/9807006 (1998).

\bibitem{Aaronson2004} S. Aaronson and  D. Gottesman,  Improved simulation of stabilizer circuits,  Phys. Rev. A {\bf70}, 052328 (2004).



 










\bibitem{BravyiHaah2012} S. Bravyi and J. Haah,  Magic-state distillation with low overhead, Phys. Rev. A {\bf86}, 052329  (2012).



\bibitem{Howard2017}M. Howard and E. T. Campbell, Application of a resource theory for magic states to fault-tolerant quantum
computing, Phys. Rev. Lett. {\bf118}, 090501 (2017).


\bibitem{WangWilde2020} X. Wang, M. M.  Wilde, and Y.  Su, Efficiently computable bounds for magic state distillation,  Phys. Rev.
Lett.  {\bf124}, 090505 (2020).

\bibitem{Regula2017} B.  Regula, Convex geometry of quantum resource
 quantification, J. Phys. A: Math. Theor. {\bf51}, 045303
 (2017).

\bibitem{ziwenliu2019} Z. W.  Liu, K. F. Bu, and R. Takagi, One-shot
 operational quantum resource theory, Phys. Rev. Lett.
 {\bf123}, 020401 (2019).

\bibitem{BravyiSmith2016}S.  Bravyi, G.  Smith, and J. A. Smolin, Trading classical and quantum computational resources, Phys.
 Rev. X {\bf6}, 021043 (2016).


 \bibitem{BravyiBrowne2019} S. Bravyi, D. Browne, P. Calpin, E. Campbell, D. Gosset, and M. Howard, Simulation of
 quantum circuits by low-rank stabilizer decompositions,
 Quantum {\bf3}, 181 (2019).



\bibitem{Dai2022} H. Dai, S. Fu, and S. Luo, Detecting magic states via
 characteristic functions, Int. J. Theor. Phys. {\bf 61}, 35
(2022).

\bibitem{feng2022} L. Feng and S. Luo, From stabilizer states to SIC-POVM fiducial states, Theor. Math. Phys. {\bf213}, 1747 (2022).


\bibitem{Veitch2014} V. Veitch, S. A. H. Mousavian, D. Gottesman, and J. Emerson, The resource theory of stabilizer quantum computation, New J. Phys. \textbf{16},  013009 (2014).

\bibitem{Souza2011}A. M. Souza, J Zhang, C. A.  Ryan, and  R. Laflamme, Experimental magic state distillation for fault-tolerant quantum computing,  Nat.  Commun. {\bf2}, 169 (2011).

\bibitem{Rodriguez2025}P. S. Rodriguez, J. M. Robinson, P. N. Jepsen, Z. He, C.
Duckering, C. Zhao, K.-H. Wu, J. Campo, K. Bagnall, M. Kwon {\sl et al}.,
  Experimental demonstration of logical magic state distillation, Nature {\bf645}, 620-625 (2025).

\bibitem{Cao2024} C. Cao, G. Cheng, A.  Hamma, L.  Leone, W.  Munizzi, and  S. F. Oliviero,   Gravitational back-reaction is magical,  arXiv:2403.07056 (2024).

\bibitem{CCao2024} C. Cao,  Non-trivial area operators require non-local magic,  J.  High Energy Phys. {\bf2024}, 1-17 (2024).

\bibitem{Andreadakis2025} F. Andreadakis and P.   Zanardi,   An exact link between nonlocal magic and operator entanglement,  arXiv:2504.09360 (2025). 

\bibitem{Ding2025} Y. M. Ding,  Z.  Wang, and Z. Yan, Evaluating many-body stabilizer Rényi entropy by sampling reduced Pauli strings: Singularities, volume law, and nonlocal magic, PRX Quantum {\bf6}, 030328 (2025).

\bibitem{WhiteCao2021} C. D. White, C. J. Cao, and  B. Swingle, Conformal field theories are magical, Phys. Rev. B {\bf103}, 075145 (2021).

\bibitem{Leone2022}L. Leone, S. F. E. Oliviero, and A. Hamma,  Stabilizer rényi entropy,  Phys. Rev. Lett. \textbf{128}, 050402 (2022).

\bibitem{Frau2024}M. Frau, P. S.  Tarabunga, M. Collura, M. Dalmonte, and E.  Tirrito,  Nonstabilizerness versus entanglement in matrix product states, Phys. Rev. B  {\bf110}, 045101 (2024).


\bibitem{Tarabunga2025}P. S. Tarabunga and  T. Haug, Efficient mutual magic and magic capacity
with matrix product states, arXiv: 2504.07230 (2025).

\bibitem{Frau2025} M. Frau, P. S. Tarabunga, M.  Collura,  E. Tirrito, and M. Dalmonte, Stabilizer disentangling of conformal field theories,  SciPost Phys.  {\bf18}, 165 (2025).




\bibitem{Qian2025} D. Qian and  J. Wang,  Quantum nonlocal nonstabilizerness, Phys. Rev. A \textbf{111}, 052443 (2025).

\bibitem{Feng2025} L. Feng and S. Luo, Entanglement and magic correlations generated by discrete beam-splitters, Ann. Phys. (Berl.) e00288 (2025).


\bibitem{Horodecki2009} R. Horodecki, P.  Horodecki, M. Horodecki, and  K. Horodecki,  Quantum entanglement, Rev. Mod. Phys. {\bf81}, 865-942 (2009).

\bibitem{Nielsen2010} M. A. Nielsen and I. L. Chuang, {\sl Quantum Computation and Quantum Information} (Cambridge University Press, Cambridge, UK, 2010).

\bibitem{Aspect1982} A. Aspect, J. Dalibard, and G.  Roger,  Experimental test of Bell's inequalities using time-varying analyzers, Phys. Rev.  Lett. {\bf49}, 1804 (1982).


\bibitem{Rowe2001} M. A. Rowe, D.  Kielpinski, V.  Meyer,  C. A. Sackett,  W. M. Itano, C. Monroe, and   D. J. Wineland,  Experimental violation of a Bell's inequality with efficient detection,  Nature {\bf409}, 791-794  (2001).

\bibitem{Ollivier2001} H. Ollivier and W. H. Zurek, Quantum discord: A measure of the quantumness of correlations, Phys. Rev.  Lett. {\bf88}, 017901 (2001).

\bibitem{Henderson2001} L. Henderson and  V. Vedral,  Classical, quantum and total correlations, J.  Phys. A {\bf34}, 6899 (2001).

\bibitem{SLUO2008} S. Luo,  Quantum discord for two-qubit systems,  Phys. Rev. A {\bf77}, 042303 (2008).


\bibitem{Loudon2000}R. Loudon, {\sl The Quantum Theory of Light}, 
(Oxford University Press, New York,  2000).




\bibitem{Fearn1987} H. Fearn and R. Loudon,  Quantum theory of the lossless beam splitter,  Opt.  Commun.  {\bf64}, 485-490 (1987).


\bibitem{Campos1989}R. A. Campos, Saleh, B. E. A. Saleh, and  M. C. Teich, Quantum-mechanical lossless beam splitter: SU (2) symmetry and photon statistics, Phys. Rev. A {\bf40}, 1371 (1989).





\bibitem{Weihs2001}G.  Weihs and A. Zeilinger,   Photon statistics at beam-splitters: An essential tool in quantum information and teleportation, in {\sl Coherence and Statistics of Photons and Atoms} {\bf262}, (2001).


\bibitem{Sasada2003} H. Sasada  and    M. Okamoto,  Transverse-mode  beamsplitter of a light beam and its application to quantum cryptography,  Phys. Rev.  A, {\bf68}, 012323 (2003).

\bibitem{Yangchang2024}L. P. Yang and Y.  Chang,  Quantum  beamsplitter as a Quantum Coherence Controller,  arXiv:2407.09791 (2024).


\bibitem{Lammerzahl1999} C. Lammerzahl and  C. J. Borde,  Atomic interferometry in gravitational fields: Influence of gravitation on the beam splitter,  Gen.  Relativ.   Gravitation, {\bf31}, 635-652 (1999).

\bibitem{Borde2004} C. J. Bord{\'e},  Quantum theory of atom-wave beam splitters and application to multidimensional atomic gravito-inertial sensor, Gen. Relativ.  Gravitation, {\bf36}, 475-502 (2004).


\bibitem{Brendel1992} J. Brendel, E.  Mohler,  and  W. Martienssen,  Experimental test of Bell's inequality for energy and time,  EPL  {\bf20}, 575  (1992).

\bibitem{Marinkovic2018} I. Marinkovi{\'c}, A.  Wallucks, R. Riedinger, S.  Hong, M.  Aspelmeyer, and  S. Gr{\"o}blacher, Optomechanical bell test, Phys. Rev.  Lett.  {\bf121},  220404 (2018).

\bibitem{Zetie2000} K. P. Zetie,  S. F.  Adams, and   R. M. Tocknell,  How does a Mach-Zehnder interferometer work?  Phys.  Educ.  {\bf35}, 46 (2000).


\bibitem{Berrada2013} T. Berrada, S. Van Frank, R.  B{\"u}cker, T.  Schumm,   J. F. Schaff,  and  J.  Schmiedmayer,  Integrated mach-zehnder interferometer for bose-einstein condensates, Nat. Commun.  {\bf4}, 2077 (2013).

\bibitem{XWang2002}X. Wang, Theorem for the beam-splitter entangler, Phys. Rev. A {\bf66}, 024303 (2002).

\bibitem{Chakhmakhchyan2018}  L. Chakhmakhchyan and  N. J. Cerf, Simulating arbitrary Gaussian circuits with linear optics, Phys.  Rev.  A {\bf98}, 062314 (2018).

\bibitem{kim2002} M. S. Kim, W.  Son,    V.  Bu{\v{z}}ek, and  P. L. Knight, Entanglement by a  beamsplitter: Nonclassicality as a prerequisite for entanglement,  Phys. Rev.  A  {\bf65}, 032323 (2002).


\bibitem{Qureshi2018} H. S. Qureshi, S. Ullah, and F. Ghafoor,  Hierarchy of quantum correlations using a linear  beamsplitter,  Sci. Rep. {\bf8}, 16288  (2018).



\bibitem{fu2020} S. Fu, S. Luo,  and  Y. Zhang,   From squeezing to Gaussian entanglement via  beamsplitters,  J. Phys.  B  {\bf53}, 085501 (2020).

\bibitem{liluoyue2025} H. Li, S.  Luo, and Y. Zhang, Detecting optical correlations via local photon subtraction,  Ann. Phys. (Berl.)  {\bf537}, 2400325 (2025).


\bibitem{kbu2023}K. Bu, W. Gu,  and A. Jaffe, Quantum entropy and central limit theorem,  Proc. Natl. Acad. Sci. U.S.A.  {\bf120}, e2304589120 (2023).


\bibitem{KBU22023}K. Bu,  W. Gu, and A. Jaffe,  Discrete quantum Gaussians and central limit theorem, arXiv:2302.08423 (2023).



\bibitem{OlivaresParis2011} S. Olivares and  M. G. A. Paris, Fidelity Matters: The birth of entanglement in the mixing of Gaussian states, Phys. Rev. Lett. {\bf107}, 170505 (2011).


\bibitem{Brunelli2015} M. Brunelli, C.  Benedetti, S. Olivares, A.  Ferraro, and M. G. A.  Paris,  Single-and two-mode quantumness at a beam splitter,  Phys.  Rev.  A {\bf91}, 062315 (2015).


\bibitem{SFuLuo2019}S. Fu, S. Luo, and Y. Zhang, Converting nonclassicality to quantum correlations via beamsplitters, EPL {\bf128}, 30003 (2019).

\bibitem{Wootters1987} W. K. Wootters, A Wigner-function formulation of finite-state quantum mechanics, Ann. Phys. \textbf{176}, 1 (1987).

\bibitem{Gibbons2004} K. S. Gibbons, M. J. Hoffman, and W. K. Wootters, Discrete phase space based on finite fields, Phys. Rev. A \textbf{70}, 062101 (2004).


\bibitem{Veitch2012} V. Veitch, C. Ferrie, D. Gross, and J. Emerson, 
Negative quasi-probability as a resource for quantum computation,
New J. Phys. \textbf{14}, 113011 (2012).



\bibitem{Gross2006} D. Gross, Hudson's theorem for finite-dimensional quantum systems, 
J. Math. Phys. \textbf{47}, 122107 (2006).













\bibitem{Wang2019} X. Wang, M. M. Wilde, and Y. Su, Quantifying the magic of quantum channels, New J. Phys. \textbf{21}, 103002 (2019).

\bibitem{Fan2003} H. Fan, H. Imai, K. Matsumoto, and X. Wang, Phase-covariant
quantum cloning of qudits, Phys. Rev. A {\bf67}, 022317 (2003).


\bibitem{Scarani2005} V. Scarani, S. Iblisdir, N. Gisin, and A. Ac{\'\i}n, Quantum cloning,
Rev. Mod. Phys. {\bf77}, 1225 (2005).



\bibitem{Alicki2004} R. Alicki and M. Fannes,  Continuity of quantum conditional information, J. Phys.  A:  Math.   Gen. {\bf37}, L55 (2004).

\bibitem{Brandao2011} F. G. Brandao, M. Christandl, and J.  Yard,  Faithful squashed entanglement,  Commun. Math. Phys.  {\bf306}, 805-830 (2011).

\bibitem{Schumacher1996} B. Schumacher and    M. A. Nielsen,  Quantum data processing and error correction, Phys. Rev.  A  {\bf54}, 2629 (1996).

\bibitem{Lloyd1997} S. Lloyd,  Capacity of the noisy quantum channel, Phys.  Rev.  A  {\bf55}, 1613 (1997).

\bibitem{Holevo2012} A. S. Holevo,  {\sl Quantum systems, channels, information: A mathematical introduction},  De Gruyter Studies in Mathematical Physics (De Gruyter Publishing, Berlin, 2012).


\bibitem{White2021} C. D. White, C. Cao, and B. Swingle, Conformal field theories are magical, Phys. Rev. B \textbf{103}, 075145 (2021).



\bibitem{Tarabunga2023} P. S. Tarabunga, E. Tirrito, T.  Chanda, and M. Dalmonte, Many-body magic via pauli-markov chains—from criticality to gauge theories, PRX Quantum {\bf4}, 040317 (2023).

\bibitem{Lopez2024}J. A. M. L{\'o}pez  and  P.  Kos,  Exact solution of long-range stabilizer Rényi entropy in the dual-unitary XXZ model,  J.  Phys.  A: Math. Theor.  {\bf57}, 475301 (2024).

\bibitem{Haug2023} T. Haug and L. Piroli, Stabilizer entropies and nonstabilizerness
 monotones, Quantum {\bf7}, 1092 (2023).


\bibitem{Sarkar2020} S. Sarkar, C. Mukhopadhyay, and A. Bayat, Characterization of an operational quantum resource in a critical many-body system, New J. Phys. \textbf{22}, 083077 (2020).


\bibitem{Fliss2021} J. R. Fliss, Knots, links, and long-range magic, J. High Energy Phys. \textbf{04}, 090 (2021).








\end{thebibliography}
\end{document}